\documentclass[final,5p,times,twocolumn]{elsarticle}

\usepackage{graphicx}

\usepackage{amssymb}
\usepackage{amsthm}

\newcounter{bla}

\journal{Computer Physics Communications}

\usepackage[english]{babel}
\usepackage[utf8]{inputenc}
\usepackage{amsmath, latexsym}
\usepackage{amsfonts}
\usepackage{physicsShortCut}
\usepackage{tdgcmNotations}
\usepackage[colorlinks=true]{hyperref}
\usepackage{url}
\usepackage{cuted}
\usepackage{dblfloatfix}

\graphicspath{{figures/}}

\begin{document}

\begin{frontmatter}



\title{\theCode-2.0: New version of the finite element solver for the time dependent generator coordinate method with the Gaussian overlap approximation.}

\author[a]{D. Regnier\corref{author}}
\author[b]{N. Dubray}
\author[b]{M. Verri\`ere}
\author[c]{N. Schunck}

\cortext[author] {Corresponding author.\\\textit{E-mail address:} regnier@ipno.in2p3.fr}
\address[a]{Institut de Physique Nucléaire, IN2P3-CNRS, Université Paris-Sud, F-91406 Orsay Cedex, France}
\address[b]{CEA, DAM, DIF, 91297 Arpajon, France}
\address[c]{Nuclear and Chemical Science Division, Lawrence Livermore National Laboratory, Livermore, CA 94551, USA}

\begin{abstract}
The time-dependent generator coordinate method (TDGCM) is a powerful method to study the large amplitude collective motion of quantum many-body systems such as atomic nuclei. Under the Gaussian Overlap Approximation (GOA), the TDGCM leads to a local, time-dependent Schr\"odinger equation in a multi-dimensional collective space. In this paper, we present the version 2.0 of the code FELIX that solves the collective Schr\"odinger equation in a finite element basis.
This new version features: (i) the ability to solve a generalized TDGCM+GOA equation with a metric term in the collective Hamiltonian, (ii) support for new kinds of finite elements and different types of quadrature to compute the discretized Hamiltonian and overlap matrices, (iii) the possibility to leverage the spectral element scheme, (iv) an explicit Krylov approximation of the time propagator for time integration instead of the implicit Crank-Nicolson method implemented in the first version, (v) an entirely redesigned workflow.
We benchmark this release on an analytic problem as well as on realistic two-dimensional calculations of the low-energy fission of \Pu240 and \Fm256. 
Low to moderate numerical precision calculations are most efficiently performed with simplex elements with a degree 2 polynomial basis. Higher precision calculations should instead use the spectral element method with a degree 4 polynomial basis. We emphasize that in a realistic calculation of fission mass distributions of \Pu240, \theCode-2.0 is about 20 times faster than its previous release (within a numerical precision of a few percents).
\end{abstract}


\end{frontmatter}



{\bf NEW VERSION PROGRAM SUMMARY}

\begin{small}
\noindent
{\em Program Title:} \theCode-2.0                                         \\
{\em Licensing provisions: GPLv3 }                                   \\
{\em Programming language:}   C++                                \\
{\em Journal reference of previous version:} \\
Computer Physics Communications 200, 350-363 (2016)    \\
{\em Does the new version supersede the previous version?:} yes  \\
{\em Reasons for the new version:}\\
\theCode-2.0 extends the physics capabilities of the previous version, since it includes the option to have a metric term in the collective Schr\"odinger equation. The computational efficiency of the code is considerably increased thanks to the implementation of several new numerical methods. This version also provides a more flexible and robust workflow to perform calculation of fission fragment distributions more efficiently.\\
{\em Summary of revisions:} \\
The new version includes the ability to use a metric term in the TDGCM+GOA equation. Numerical integration methods (both spatial and time integration) have been updated to more efficient schemes: matrix elements are now computed with Gauss-Legendre or Gauss-Legendre-Lobatto quadratures, the time propagation of the wave packet is computed with the Krylov approximation of the propagator instead of the previous Crank-Nicolson scheme. A new type of finite element based on n-dimensional orthotope is implemented, which enables a spectral element scheme for spatial discretization. Finally, the workflow and inputs/outputs of the code were entirely redesigned to provide more flexibility and be more user-friendly.
\\
{\em Nature of problem:} \\
The Gaussian overlap approximation to the time-dependent generator coordinate method [1,2] yields a local, time-dependent Schr\"odinger equation in a small, multi-dimensional collective space. Its solution provides the time-evolution of the collective probability amplitude. For applications to nuclear fission, scission configurations are defined by an hyper-surface in this collective space referred to as the frontier. Distributions of nuclear observables such as charge or mass distributions of the fission fragments are extracted from the probability for the system to pass through any given section of the frontier. This probability is computed by first solving the evolution equation up to times greater than $10^{-20}$s and then integrating the flux of probability through the frontier over the entire time range. 
\\
{\em Solution method:} \\
\theCode-2.0 solves the time-dependent Schr\"odinger equation by first discretizing the N-dimensional collective space with the continuous Galerkin Finite Element Method. This produces a large set of coupled, time-dependent Schr\"odinger equations characterized by the sparse overlap and Hamiltonian matrices. The solution is evolved in small time steps by applying an explicit and unitary propagator built as a Krylov approximation [3] of the exponential of the Hamiltonian.\\
{\em Additional comments including Restrictions and Unusual features:}\\
Although the implementation of the program gives it the ability to solve the TDGCM+GOA equation in a generic N-dimensional collective space, it has only been tested on 1-, 2- and 3-dimensional meshes.
   \\

\end{small}

\section{Introduction}
\label{sec:introduction}
A completely microscopic description of the fission process is a major challenge for nuclear theory. Fission is a time-dependent, non-equilibrium, quantum many-body problem where more than 200 nucleons interact over large time scales $\tau$ (typically $\tau > 10^{-20}$s) in a coherent way to break the system into two or more fragments. Of particular interest to applications of fission in either science (nucleosynthesis, superheavy science) or applications (energy production) are the properties of the fission fragments, in particular their charge and mass distributions. One of the most effective approaches to computing such observables in a quantum-mechanical framework is the Gaussian overlap approximation (GOA) of the time-dependent generator coordinate method (TDGCM)~\cite{griffin_collective_1957,reinhard_generator-coordinate_1987}. This approach relies on the energy density functional (EDF) formalism and reduces the dynamics of the complete system to a local, collective Schr\"odinger-like equation involving only a few arbitrary degrees of freedom referred to as collective variables. This approach was originally introduced for the description of low-energy neutron-induced fission in the 1980ies \cite{berger_microscopic_1984,berger1986,berger_time-dependent_1991} but was not applied on a large scale because of computational limitations at the time. Thanks to progress in computing capabilities, the TDGCM+GOA approach to fission has been recently applied to predictions of fission fragment distributions in 2-dimensional collective space\cite{goutte_microscopic_2005,younes_fragment_2012,younes_collective_2012,regnier_fission_2016,zdeb_fission_2017}. These studies, together with others~\cite{dubray_numerical_2012}, have highlighted the need to take into account additional degrees of freedom in order to increase the accuracy of calculations. The first version of \theCode \ was a first step towards the goal of providing the community with a numerical tool capable of solving the TDGCM+GOA equation for an arbitrary number of collective variables \cite{regnier_felix-1.0:_2016}. In this paper, we present a major upgrade of \theCode \ which contains both new physics features and much improved numerical performances.

In section~\ref{sec:modifications}, we review the new and upgraded features of \theCode-2.0 compared to the previous version. The convergence properties of the different numerical methods implemented are benchmarked in section ~\ref{sec:benchmarks}. We also present in the same section a comparison of the performances between \theCode-2.0 and \theCode-1.0. The section~\ref{sec:usage} is finally devoted to the practical installation and usage of the package.

\section{Modifications introduced in version 2.0}
\label{sec:modifications}

\subsection{The TDGCM+GOA with a metric}

The time-dependent extension to the generator coordinate method provides an appropriate formalism to describe the slow and large amplitude motion of
nuclei \cite{reinhard_generator-coordinate_1987,ring2000}. In this approach, we assume that the many-body state $| \Psi(t) \rangle$ of the fissioning system takes the generic form
\begin{equation}
\label{eq:gcmApprox}
|\Psi(t)\rangle= \int_{\qvec} f(\qvec, t) |\Phi_{\qvec} \rangle \, \text{d}\qvec.
\end{equation}
The set $\{ |\Phi_{\qvec}\rangle \}_\qvec$ is made of known many-body states parametrized by a vector of continuous variables $\qvec \equiv (q_{1}, \dots, q_{N})$. Each of these $q_{i}$ is a collective variable and must be chosen based on the physics of the problem. Inserting the ansatz (\ref{eq:gcmApprox}) in the time-dependent many-body Schr\"odinger equation for the fissioning nuclei, yields an equation for the unknown weight function $f(\qvec, t)$, the Hill-Wheeler equation,
\begin{equation}
 \forall \,\qvec: \quad \int_{\qpvec} \langle \Phi_{\qvec} | \left[ \hat{H} - i\hbar \frac{\partial }{\partial t}\right]|  \Phi_{\qpvec} \rangle f(\qpvec,t) \,\text{d}\qpvec = 0.
\end{equation}
In this work, we will assume that the potential part of the nuclear Hamiltonian $\hat{H}$ is approximated by an effective two-body interaction of the Skyrme or Gogny type. 
Numerically solving the time-dependent Hill-Wheeler equation is a difficult task requiring a tremendous amount of computational resources. Although such a direct approach is currently being investigated~\cite{verriere_fission_2017}, a widespread alternative consists in injecting an additional assumption about the generator states $| \Phi_{\qvec} \rangle$, known as the Gaussian overlap approximation (GOA)~\cite{reinhard_generator-coordinate_1987,brink1968,krappe2012}. In its simplest formulation, the GOA assumes that the overlap between two generator states $\langle \Phi_\qvec | \Phi_\qpvec \rangle$ has a Gaussian shape that depends on the difference $(\qvec-\qpvec)$. With this new version of the code, we make it possible to use a more flexible version of this method introduced in~\cite{kamlah_derivation_1973,onishi1975,god1985-a}. We assume that one can find a change of variables $\qvec\rightarrow\boldsymbol{\alpha} = \boldsymbol{\alpha}(\qvec)$ such that the overlap reads
\begin{equation}
\label{eq:goa}
\langle \Phi_\qvec | \Phi_\qpvec \rangle \simeq
\operatorname{exp}\left( -\frac{1}{2} \sum_k \left[ \alpha_k(\qvec)- \alpha_k(\qpvec)\right]^2 \right).
\end{equation}
Within this approximation, the Hill-Wheeler equation reduces to a local, time-dependent Schr\"odinger-like equation in the space $\mathcal{Q}$ of the coordinates $\qvec$,
\begin{equation}
\label{eq:tdgcmgoa}
i\hbar \frac{\partial g(\qvec,t)}{\partial t} = \hat{H}_\text{coll}(\qvec) \, g(\qvec,t).
\end{equation}
The complex function $g(\qvec,t)$ is the unknown of the equation. It is related to the weight function $f(\qvec, t)$ appearing in (\ref{eq:gcmApprox}) and contains all the information about the dynamics of the system. The collective Hamiltonian $\hat{H}_\text{coll}(\qvec)$ is a local operator acting on $g(\qvec,t)$,
\begin{equation}
\label{eq:Hcoll}
\hat{H}_\text{coll}(\qvec)=
-\frac{\hbar ^2}{2\gamma^{1/2}(\qvec)} \sum_{ij} \frac{\partial}{\partial q_i} \gamma^{1/2}(\qvec) B_{ij}(\qvec) \frac{\partial}{\partial q_j}  +  V(\qvec),
\end{equation}
where
\begin{itemize}
\item The potential $V(\qvec)$ and the symmetric, collective inertia tensor $\Bvec(\qvec) \equiv B_{ij}(\qvec)$ can be determined from the original nuclear Hamiltonian $\hat{H}$ and the generator states $|\Phi_{\qvec} \rangle $. These quantities reflect the potential and kinetic properties of the system in the collective space.
\item The metric $\gamma(\qvec)$ is a positive, real, scalar field introduced by the change of variable $\qvec\rightarrow\boldsymbol{\alpha}(\qvec)$ used for the GOA approximation.
\end{itemize}
Equation~ (\ref{eq:tdgcmgoa}) will be referred to as the TDGCM+GOA equation. Compared to \theCode-1.0, this version solves the TDGCM+GOA equation with an arbitrary real positive metric field $\gamma(\qvec)$. To take advantage of this option, the user must provide the value $\gamma(\qvec)$ of the metric field at each point of the mesh alongside the collective inertia tensor $\Bvec(\qvec)$ and the potential energy $V(\qvec)$. While the code accepts any metric, the GCM metric is the most natural and consistent choice; see \cite{schunck_microscopic_2016} and references therein for details.

\subsection{Improved numerical methods}

\subsubsection{Support for a new type of finite element}
\label{sec:elements}

The first step toward solving (\ref{eq:tdgcmgoa}) consists in discretizing the collective space into a finite element basis. To do so, we first partition the simulation domain $\Omega$ of the collective space into a mesh of small cells. We also define a set of specific points $\{\qvec_i \ |\  i \in [1,m]\}$ called nodes. A subset of nodes $\{\qvec_i \ | \ i\in \mathcal{N}_c\}$ is assigned to each cell $c$. The numerical solution of (\ref{eq:tdgcmgoa}) is sought as a continuous function that takes a polynomial shape into each cell of the mesh. We note $\mathcal{P}_c$ the vector space of all allowed interpolating polynomials in a cell $c$. It is defined in terms of the Lagrange polynomials built on  the nodes $\{\qvec_{i}\ | i \in \mathcal{N}_c \}$,
\begin{equation}
 \mathcal{P}_c = \text{span}\{l_{c,i} \ | i \in \mathcal{N}_c \} 
\end{equation}
where $l_{c,i}$ is the Lagrange polynomial whose value is one at $\qvec_{i}$ and zero at the other nodes of the cell $c$.
This property yields a convenient expansion of the numerical solution in the finite element basis,
\begin{equation}
\label{eq:gExpand}
g(\qvec,t)= \sum_{ i=1}^m g(\qvec_i,t) \phi_i(\qvec)
\end{equation}
The basis functions $\{\phi_{i}\}_i$ are defined as
\begin{equation} 
\forall \qvec \in \Omega: \quad \phi_{i}(\qvec)= \left \{ 
\begin{array}{ll} 
l_{c,i}(\qvec), & \text{if } \qvec \in c \text{ and } i\in \mathcal{N}_c, \\ 
0, & \text{otherwise}. \\ 
\end{array} \right. 
\end{equation} 
Guaranteeing the continuity of the solution imposes constraints on the cells and nodes. In particular, the maximum degree of the polynomials in $\mathcal{P}_c$ must be the same for each cell. Also, two adjacent cells must share a sufficient number of nodes at their interface. 

This version of FELIX supports two kinds of cells: the n-dimensional simplex (already available in version 1.0) and the new $N$-orthotope (or n-dimensional hyper-rectangles).
A n-dimensional simplex generalizes the notion of triangle to an arbitrary dimension. It is formally defined as the convex hull of $N$-vectors being a basis of the $N$-dimensional space (triangle if $N=2$, tetrahedron if $N=3$, etc.). In such a cell, we define the space $\mathcal{P}_{c}$ as the set of polynomials generated by any linear combination of monomials with a total degree lower than $ d_c$,
\begin{equation}
\mathcal{P}_c = \text{span} \left\{
\displaystyle 
m(q_1,\cdots, q_N) = \prod_{i=1}^{N} q_i ^{d_i}\ \left| \ \sum_{i = 1}^{N} d_i \le d_c \right. 
\right\} 
\end{equation}
where the total degree $ d_c $ is chosen by the user.
The dimension of this space is given by the binomial coefficient,
\begin{equation}
\text{dim}(\mathcal{P}_c) = \left( \begin{array}{c} N+d_{c} \\ d_{c} \end{array}\right).
\end{equation}
A $N$-orthotope generalizes the notion of a rectangle to an arbitrary dimension. It can be defined as a Cartesian product of one one-dimensional interval for each dimension. The edges of the $N$-orthotope may have different lengths but no other deformation or rotation is authorized. For this kind of cells, we use a different interpolating polynomial set than for the simplex cell. The space $\mathcal{P}_{c}$ is spanned by any linear combination of monomials with all exponents less than or equal to the degree $ d_c $,
\begin{equation}
\mathcal{P}_c = \text{span} \left\{
\displaystyle 
m(q_1,\cdots, q_N) = \prod_{i=1}^{N} q_i ^{d_i}\ \left| \ \forall i, \ d_i \le d_c \right. 
\right\} 
\end{equation}
The maximum degree $ d_c $ is chosen by the user.
The dimension of this space is simply given by
\begin{equation}
\text{dim}(\mathcal{P}_c) = (d_c + 1)^{N} 
\end{equation}

The finite element basis is entirely determined by the geometry of the cells together with the number and positions of the nodes. The setup tool provided with \theCode\ (cf. section~\ref{sec:usage}) is capable of automatically generating a consistent finite element basis from a set of user-defined basis parameters. The supported types of finite element basis are summarized in Table~\ref{tab:finiteElements}. 

\begin{table}[!ht]
\begin{center}
 \begin{tabular}{ccc}
 \hline
 Cell & Degree ($d_c$) & nodes position \\
 \hline
 \textit{simplex} & 1,2         & regular\\
 \textit{hcube}   & any & Gauss-Lobatto\\
 \hline
 \end{tabular}
 \caption{Finite element bases that can be generated by the setup tool.}
 \label{tab:finiteElements}
 \end{center}
\end{table} 

\begin{figure}[!ht]
\begin{center}
\includegraphics[width=0.45\textwidth]{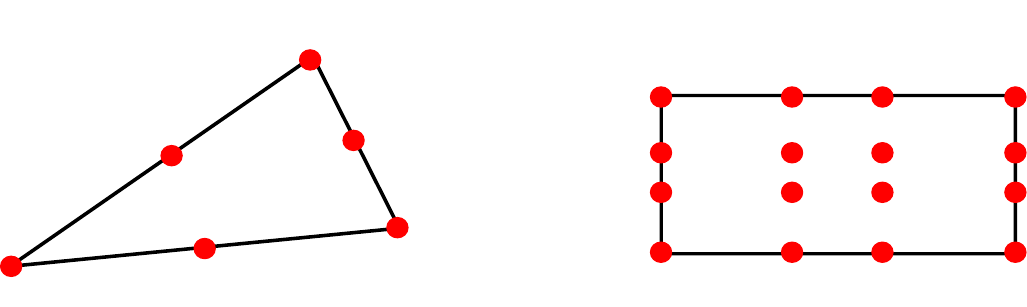} 
\caption{Examples of cells with their nodes (red points) defining the element basis. Left: 2-simplex of degree 2; The nodes are regularly spaced in the basis generated by two edges of the cell. Right: 2-orthotope of degree 3; The nodes are positioned at the Gauss-Lobatto quadrature points of the cell.}
\label{fig:elementExample}
\end{center}
\end{figure}

The keyword \textit{hcube} refers to $N$-orthotope cells.
The two possible schemes for node positioning are
\begin{itemize}
 \item regular: The nodes are regularly spaced in the intrinsic basis of the cell.
 \item Gauss-Lobatto: The nodes are located at the Gauss-Lobatto quadrature points of the cell.
\end{itemize}
Two typical cell examples are illustrated in figure~\ref{fig:elementExample}.

\subsubsection{Computation of matrix elements with quadratures}

After spatial discretization, the TDGCM+GOA equation turns into
\begin{equation} 
\label{eq:spaceDiscretizedEvolution}
i\hbar M \, \frac{\partial \textbf{g}(t)}{\partial t}= [H -i\hbar A] \textbf{g}(t) 
\end{equation}
The vector $\textbf{g}(t)$ has length $m$ and contains the unknown coefficients $g(\qvec_i,t)$ at every node $i$. The $m\times m$ matrices $ M, H$ and $A$ are defined as
\begin{equation}
\label{eq:matrixElements1}
\begin{array}{rl} 
M_{ab} = & \displaystyle \int_\Omega d\qvec\; \phi_a(\qvec)\, \left[\sqrt{\gamma(\qvec)}\right] \,\phi_b(\qvec), \medskip \\
A_{ab} = & \displaystyle \int_\Omega d\qvec\; \phi_a(\qvec)\, \left[A(\qvec)\,\sqrt{\gamma(\qvec)}\right] \,\phi_b(\qvec), \medskip \\
H_{ab}=  & \displaystyle \int_\Omega d\qvec\; \phi_a(\qvec)\, \left[V(\qvec)\,\sqrt{\gamma(\qvec)}\right] \,\phi_b(\qvec) \medskip \\
         & +\displaystyle \frac{\hbar^2}{2} \sum_{kl}  \int_\Omega d\qvec\;  \frac{\partial \phi_a}{\partial q_k}(\qvec) \left[B_{kl}(\qvec)\sqrt{\gamma(\qvec)}\right] \frac{\partial \phi_b}{\partial q_l} (\qvec). \\
\end{array} 
\end{equation}
The field $A(\qvec)$ is provided by the user and enables an absorption of the collective wave packet close to the boundary of the simulation domain.
The matrix element associated with the kinetic part is obtained by integrating by parts the double derivative in space and using the Dirichlet conditions imposed on the boundary $\partial \Omega$. The general approach used to compute efficiently these matrix elements consists in assuming that the integrand are polynomials. In the previous release of \theCode, these integrals were determined by the computation of the symbolic expression of the integrated polynomial followed by the evaluation of this expression where needed. In the new version, the integrals are evaluated thanks to an adapted quadrature with a sufficiently high order. This method typically speeds up the calculation of the Hamiltonian matrix by a factor 100 for a realistic 2-dimensional calculation on \Fm256.

\begin{figure}[!ht]
\begin{center}
\includegraphics[width=0.45\textwidth]{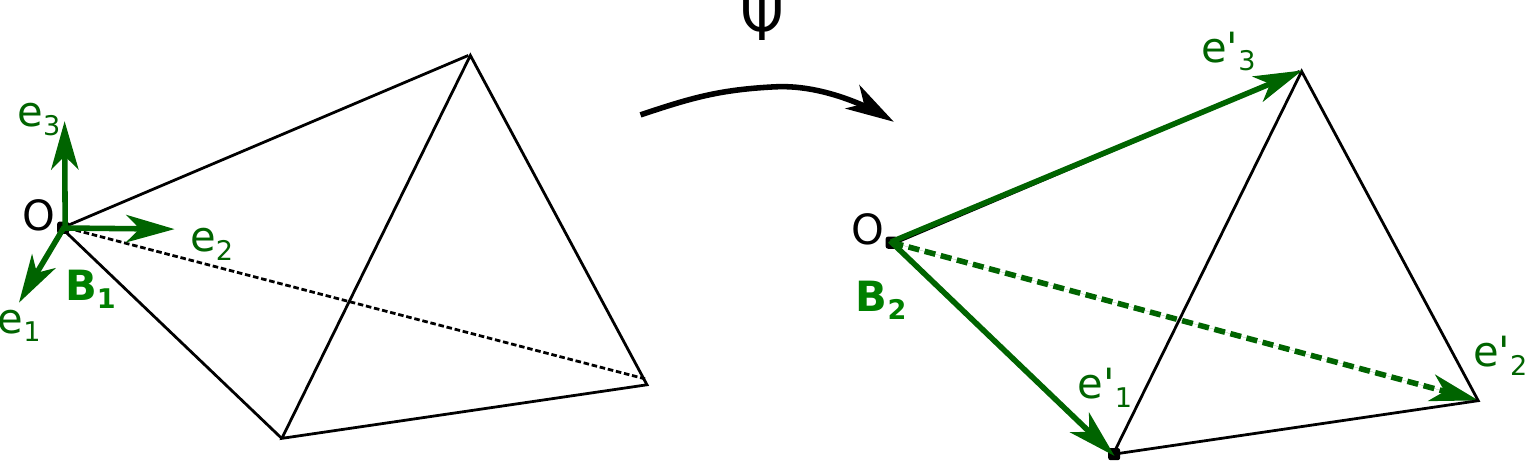} 
\caption{Change of coordinate $\psi$ in the case of a 3-dimensional simplex cell. $B_1$ is the direct 3-dimensional canonical frame centered on a vertex $O$ of the cell. $B_2$ is the intrinsic frame of the cell centered on $O$ and corresponds to the $\psi$ transformation of $B_1$.}
\label{fig:intSimplex}
\end{center}
\end{figure}

In practice, we assume that any of the fields listed above can be expanded in the finite element basis
\begin{equation}
\label{eq:fieldExpand}
F(\qvec)= \displaystyle\sum_{c=1}^m F(\qvec_c)\,\phi_c(\qvec) \quad \text{with } F=\sqrt{\gamma},\, A\sqrt{\gamma}, \dots
\end{equation}
The integrals in (\ref{eq:matrixElements1}) are then expressed as sums of integrals over the $ n_c $ cells $ c_i $ of the mesh,
\begin{align}
\label{eq:matrixElements}
M_{ab} = & \displaystyle \sum_{i=1}^{n_c} \sum_{c=1}^m \sqrt{\gamma(\qvec_c)} \int_{c_i}d\qvec\; \phi_a(\qvec)\,\phi_b(\qvec)\,\phi_c(\qvec), \medskip \nonumber \\
A_{ab}= & \displaystyle \sum_{i=1}^{n_c} \sum_{c=1}^m  A(\qvec_c) \sqrt{\gamma(\qvec_c)} \int_{c_i} d\qvec\;  \phi_a(\qvec)\,\phi_b(\qvec)\,\phi_c(\qvec), \medskip \nonumber \\
H_{ab}= & \displaystyle \sum_{i=1}^{n_c} \sum_{c=1}^m  V(\qvec_c) \sqrt{\gamma(\qvec_c)}\int_{c_i} d\qvec\;  \phi_a(\qvec)\,\phi_b(\qvec)\,\phi_c(\qvec) \nonumber \\
        & +\displaystyle \frac{\hbar^2}{2} \sum_{i=1}^{n_c} \sum_{c=1}^m \sum_{kl}  B_{kl}(\qvec_c) \sqrt{\gamma(\qvec_c)} \nonumber \\
        & \displaystyle \times \int_{c_i} d\qvec\; \frac{\partial \phi_a}{\partial q_k}(\qvec) \frac{\partial \phi_b}{\partial q_l} (\qvec)\, \phi_c(\qvec).
\end{align}
Any integral over a cell is computed in two steps. First we apply a change of coordinate $ \psi $ in order to transform the domain of integration into a reference domain. The reference domain for a $N$-simplex is the $N$-simplex generated by the canonical orthonormal basis. The reference domain for a $N$-orthotope is a $N$-hypercube with edges of length 1. This change of coordinate $\psi$ can also be viewed as the linear application that transforms the canonical basis $B_1$ of the N-dimensional space into an intrinsic basis $B_2$ of the cell, whose vectors lie on the edge of the cell (see Fig.~\ref{fig:intSimplex} for an illustration on the 3-dimensional simplex cell).
Using this transformation, we can recast our integrals into equation (\ref{eq:matrixIntegral}), where $J_{\psi}$ stands for the Jacobian matrix of the transformation $\psi$.

\begin{table*}[!ht]
\begin{equation}
\label{eq:matrixIntegral}
\begin{array}{rcll}
\displaystyle \int_{c_i} d\qvec\;  \phi_a(\qvec)\,\phi_b(\qvec)\,\phi_c(\qvec) & = &
\displaystyle \int_{q'_1=0}^{1} \int_{q'_2=0}^{1-q'_1}\cdots\int_{q'_N=0}^{1-q'_1-q'_2\cdots-q'_{N-1}} d\qvec' |\text{det}(J_\psi)| \; (\phi_a\phi_b\phi_c)\circ\psi(q'_1,\cdots,q'_N), & \text{for a simplex}, \\
\\
  & = & \displaystyle \int_{q'_1=0}^{1} \int_{q'_2=0}^{1}\cdots\int_{q'_N=0}^{1} d\qvec' |\text{det}(J_\psi)|\; (\phi_a\phi_b\phi_c)\circ\psi(q'_1,\cdots,q'_N), & \text{for an orthotope}.
\end{array}
\end{equation}
\end{table*}

For the special case of the kinetic term, some basis polynomials are replaced with their derivatives but the  derivation remains the same.
In a second step, each 1-dimensional integral is computed by quadrature. The type of quadrature is defined by the user and the current version of the code implements Gauss-Legendre and Gauss-Lobatto quadratures. 
The same order of quadrature $ n_q $ is used in each dimension.
We recall that exact integration of a degree $d$ polynomial is achieved if
\begin{equation}
\begin{array}{ll}
d \le 2n_q -1 & \text{Gauss-Legendre}, \\
d \le 2n_q -3 & \text{Gauss-Lobatto}.\\
\end{array}
\end{equation}
The Gauss-Legendre quadrature is therefore more efficient but only involves nodes positioned inside the integration domain. On the other hand, the Gauss-Lobatto quadrature includes the boundary of the integration domain as nodes. As discussed in section~\ref{sec:spectralelem}, this property is crucial for the spectral element discretization scheme.
The highest polynomial exponent $ d $ to be integrated with the quadrature during the full multi-dimensional integration verifies
\begin{equation}
\begin{array}{ll}
d \le 3 d_{c_i} & \text{for a $N$-orthotope} \\
d \le 3 N d_{c_i} + N-1 & \text{for a $N$-simplex} \\
\end{array}
\end{equation}
In practice, the code determines the order $ n_q $ needed to perform integrations exactly. 
For instance, the matrix elements of a $N$-simplex with $ d_{c_i}=2 $ will be computed with $ n_q = \frac{1}{2}(7N+3 - (7N+1)\%2)$, if a Gauss-Legendre quadrature is adopted ($\%2$ denotes the rest in the Euclidian division by 2).

\subsubsection{The spectral element scheme}
\label{sec:spectralelem}

The general approach described above produces the spatially discretized equation~(\ref{eq:spaceDiscretizedEvolution}) with a non trivial overlap matrix $M$. As explained in section~\ref{sec:krylov}, the time integration of this equation requires inverting the matrix $M$ at each time step. The spectral element method consists in choosing a specific set of cells type, node positions and quadrature that leads to an easily invertible, diagonal $M$ matrix. The main idea is to trade some precision in the evaluation of the matrix elements to accelerate time propagation.

To build the spectral element basis, we first require that all cells of the mesh are $N$-orthotopes. The user defines the degree $d_c $ of the polynomial interpolating functions for each cell. Then the nodes of each cell should be located on the grid generated by the zeros of the $n_q = d_c +1 $ order Gauss-Lobatto polynomials in each dimension. This corresponds to the \textit{hcube} elements described in section~\ref{sec:elements}. Because the Gauss-Lobatto quadrature includes the integration domain boundaries as nodes, it is still possible to ensure the continuity of the finite element trial solution (\ref{eq:gExpand}).  In this configuration, the finite element basis function $ \phi_a $ obey
\begin{equation}
\phi_a(\qvec_b) = \delta_{ab},
\end{equation}
for all nodes $\qvec_b$ of the $N$-dimensional Gauss-Lobatto quadrature in the cell.

The matrix elements are then approximated with the Gauss-Lobatto quadrature of order $ n_q $. Note that this order does not guarantee exact integration anymore as in the general case. On the other hand, the matrix elements of any local field $ F(\qvec) $ are diagonal and read
\begin{equation}
\begin{array}{rl}
F_{ab} = & \displaystyle\int_\Omega d\qvec\; \phi_a(\qvec)\; F(\qvec)\; \phi_b(\qvec), \\
       \\
       = & \displaystyle\sum_{i=1,\,  a\in \mathcal{N}_{c_i}}^{n_c} \sum_{c=1}^{n_q^N} |\text{det}(J_\psi^i)| w_c^i \delta_{ac} F(\qvec_c) \delta_{bc},  \\
       \\
       = & \displaystyle \delta_{ab} F(\qvec_a) \sum_{i=1,\,  a\in \mathcal{N}_{c_i}}^{n_c}  |\text{det}(J_\psi^i)| w_a^i, 
\end{array}
\end{equation}
where $i$ runs over the cells of the mesh, $J_\psi^i$ is the Jacobian of the geometric transformation of the cell $i$, and $w_a^i$ is the quadrature weight associated with the node $a$ for cell $i$.

\subsubsection{Krylov method for time propagation}
\label{sec:krylov}

While the previous version of \theCode\ relied on the implicit Crank-Nicolson scheme for time integration, we adopt here a Krylov approximation of the exact time propagator. This type of propagator is explicit (time propagation not based on a system inversion at each time step) and unitary (conservation of the norm at each time step).

Starting from (\ref{eq:spaceDiscretizedEvolution}), we first factorize the $M$ matrix,
\begin{equation}
 M = P^t L L^t P.
\end{equation}
The matrix $P$ is just a convenient reordering of the matrix index, and $L$ is the lower part of the Cholesky decomposition of $M$ after reordering. The reordering step aims at optimizing the sparsity of the Cholesky factor $L$. We define $\mathbf{\tilde{g}} = L^t P \mathbf{g}$ so that the time evolution reduces to
\begin{equation}
i \bar{h} \frac{\partial \mathbf{\tilde{g}}}{\partial t} = L^{-1} P [H - i \bar{h} A] P^t (L^t)^{-1} \mathbf{\tilde{g}} ,
\end{equation}
and, therefore,
\begin{equation}
\mathbf{\tilde{g}}(t+\Delta t) = \text{exp} \left( - \frac{i \Delta t}{\bar{h}} \tilde{H} \right) \mathbf{\tilde{g}}(t) 
\end{equation}
with $\tilde{H} = L^{-1} P [H - i \bar{h} A] P^t (L^t)^{-1} $.
To compute the exponential, the matrix $ \tilde{H}$ is approximated using the Arnoldi algorithm~\cite{saad_iterative_2003} with the the Krylov space spanned by 
$
 \{ \mathbf{\tilde{g}}(t), \tilde{H} \mathbf{\tilde{g}}(t), \tilde{H}^2 \mathbf{\tilde{g}}(t), \dots, \tilde{H}^{(n_K-1)} \mathbf{\tilde{g}}(t) \}:
$
\begin{equation}
\tilde{H} \simeq Q \tilde{h} Q^\dagger,
\end{equation}
where both  $Q $ and  $\tilde{h} $ result from the Arnoldi algorithm. The columns of the  $ \text{dim}(\mathbf{g}) \times n_K $ matrix  $Q $ contain an orthonormal basis of the Krylov space, whereas  $ \tilde{h}  $ is a Hessenberg matrix of dimension  $n_K \times n_K $. The dimension  $ n_K $ of the Krylov space is in practice selected by the user. Note that the first column of  $Q $ is by construction proportional to  $ \mathbf{\tilde{g}}(t)  $.
Accordingly, the action of the exponential propagator on $\mathbf{\tilde{g}}(t)$ reads:
\begin{align}
\text{exp} \left( - \frac{i \Delta t}{\bar{h}} \tilde{H} \right) \mathbf{\tilde{g}}(t)
 & \simeq \ 
Q \, \text{exp} \left( - \frac{i \Delta t}{\bar{h}} \tilde{h} \right) Q^\dagger \mathbf{\tilde{g}}(t), \nonumber \\
 & = \ 
||\mathbf{\tilde{g}}(t)|| \, Q \,  \text{exp}  \left( -\frac{i \Delta t}{\bar{h}} \tilde{h} \right) \bf{e_0}
\end{align}
Using a small dimension for the Krylov space (typically around 10), the exponential of the matrix $ \tilde{h} $ can be quickly computed based on a linear algebra library. Most of the numerical cost of this method resides in the application of the Arnoldi algorithm at each time step. 

In the case where the matrix $ M $ is diagonal (\textit{e.g.} usage of a spectral element space scheme), its Cholesky decomposition and the numerical inversion of $L$ and $ L^t$ become trivial. Time iterations then only involve matrix/vector multiplications and vector/vector dot products.

\subsection{Automated Workflow}

Along with the TDGCM+GOA solver itself, \theCode-2.0 comes with a set of tools that help the user manage the pre/post-treatment operations in an automated way. These tools were mostly designed with the purpose of computing fission yields in the $(Q_{20},Q_{30})$ collective space and we describe their logic in this section. To use the solver in a different context, users are encouraged to adapt/re-write these tools based on the FELIX library and the present pre/post-treatment programs. Within our automated workflow, a typical fission calculation requires three major steps: setup, solver, post-processing. Each step is associated with one executable, namely {\tt flx-setup}, {\tt flx-solver} and {\tt flx-post}.

\subsubsection{Setup (flx-setup)}
This tool takes as main input an unstructured list of points along with the values of a series of fields at these points (potential, inertia, metric, etc). Starting from this initial data, the main goal of the setup is to generate the simulation domain, the finite element basis and the input fields in a format compatible with the time evolution. This step includes both purely numerical operations, such as Delaunay triangulation of the original set of points, and transformations of the physical inputs (extrapolation of the potential in the low $Q_{20}$ region, cropping of the fusion valley, etc.). It requires the knowledge of average value of the neck operator $\hat{Q}_\text{N}(\qvec)$ as a function of the deformation. A commonly adopted definition of this operator is given in~\cite{younes_microscopic_2009} and involves a convolution of the nuclear density with a Gaussian centered at the point of minimal radial density and with a width of 1 fm. In this workflow, the value of the neck operator is provided as an input along with the potential and inertia tensor and could therefore be defined differently by the user.
The {\tt flx-setup} executable performs the following operations:
\begin{enumerate}
 \item remove the points with a neck value $\langle\hat{Q}_{N}\rangle$ lower than a user-defined threshold. This is used to keep only the fission valley;
 \item perform a Delaunay triangulation of the remaining unstructured list of points. Note that the fission valley domain is not necessarily a convex area whereas a raw Delaunay triangulation results in a convex shape. To circumvent this issue, we define the mesh as an alpha-shape~\cite{edelsbrunner_three-dimensional_1994} with a user-defined alpha parameter. A particular attention must be paid to the alpha parameter to obtain the expected fission valley.
 \item generate a degree 1 finite element basis for this mesh of simplices; 
 \item build a new user-defined mesh $M_1$ on a hyper-rectangle containing the simulation domain;
 \item copy (interpolate) the original fields (fission valley only) on the new mesh; 
 \item extrapolate the fields over the rest of the mesh $M_1$; The fields are extrapolated using an empirical formula to ensure continuity at the boundary of the original field (fission valley only). For any point $\qvec$ in the extrapolated region, the field $F(\qvec)$ is computed as
 \begin{equation}
 \label{eq:extrapol}
  F(\qvec) = \frac{1}{N } \sum_{i\in \partial \Omega_0} \frac{1}{d_i^5} F(\qvec_i),\quad  N = \sum_{i\in \partial \Omega_0} \frac{1}{d_i^5},
 \end{equation}
 where the sum runs over every node $\qvec_i$ lying on the boundary $\partial \Omega_0$ of the fission valley domain and $d_i$ is the distance separating $\qvec$ from the node $\qvec_i$. In the specific case of the potential $V(\qvec)$, an additional term $-\alpha d$ is added to the right hand side of Eq.~\ref{eq:extrapol} with $d$ being the minimal distance between $\qvec$ and the fission valley and $\alpha$ a user-defined slope to control this artificial decrease of the potential in the extrapolated region.
 \item generate an absorption field;
 The absorption field consists in a scalar field $A(\qvec)$ defined over the final domain. It is set to zero in the fission valley area (defined by $\langle\hat{Q}_N\rangle > Q_{\mathrm{cut}}$). In the outer zone, it reads
\begin{equation}
\label{eq:absorption}
A(\qvec)= 4r\left(1-\frac{w-x(\qvec)}{w}\right)^3.
\end{equation}
The parameters $r$ and $w$ define the amplitude and characteristic width of the absorption, whereas $x(\qvec)$ is the minimal Euclidian distance between $\qvec$ and the fission valley.
 \item perform h-refinement operations. This is done automatically near the ground-state of the potential well;
 \item crop the mesh $M_1$ based on the distance to the fission valley;
 \item perform p-refinement operations;
 \item symmetrize the mesh relative to the Oy axis ($Q_{30}$ symmetry);
 \item write the final finite element basis and the fields.
\end{enumerate}

Upon demand, the setup tool can also generate a set of quasi-bound states to be used by the solver to build an initial collective state for the dynamics. To do so, {\tt flx-setup} will first compute a new potential  $V'(\qvec) $ in the vicinity of a user-defined potential well $V(\qvec) $. By construction, the new potential verifies $ V'(\qvec) = V(\qvec) $ for values of $V$ lower than the first saddle point of the selected potential well. At points  $\qvec $ corresponding to larger values of  $V(\qvec) $, the new potential is determined by a degree 2 polynomial with respect to the distance $d(\qvec, q_{gs})$ to the ground state  $V'(\qvec) = a d^2(\qvec, q_{\mathrm{gs}}) + b d(\qvec, q_{\mathrm{gs}}) + c  $. The coefficients of this polynomial are tuned to match the original potential at three points of the well (the ground state, the saddle point lying on the segment $[\qvec_{gs}, \qvec] $, and the middle point between these two). Once the new potential $V'$ is determined, the setup tool compute the lowest eigenstates $g_k(\qvec)$ of the new Hamiltonian
\begin{equation}
\hat{H}'(\qvec) = \TCollExpr + V'(\qvec).
\end{equation}
The results obtained are written on disk and can be read by the solver to compute initial states of the form
\begin{equation}
g(\qvec, t=0) \propto \sum_k \text{exp}\left( \frac{E_k - E_0}{\sigma^2} \right) g_k(\qvec)
\end{equation}

\subsubsection{Solver (flx-solver)}
The role of the solver consists in building or reading an initial state and then propagating it in time. It takes as inputs the finite element basis and fields produced by {\tt flx-setup} as well as optional data relative to the initial state. Once every 'dump' time iteration, the solver will append the real and imaginary part of the collective wave function $g(\qvec, t)$ to an output file {\tt dyn.dat}. This file stores the information in binary format for the sake of numerical precision as well as memory saving. No physical observable is computed ``on the fly'' during the time evolution.

\subsubsection{Post-treatment (flx-post)}
This tool takes as main input the result of a calculation performed with {\tt flx-solver} and post-processes the data to provide human-readable information about the dynamics. It may return various results mainly depending on the {\tt action} options passed by the user; see the HTML documentation accompanying the code. If the value of {\tt action} is set to {\tt all}, post-processing involves the following steps:
\begin{itemize}
\item compute an oriented frontier (hyper-surface) as an iso-value of the field 'qN' (neck operator) (more information on frontier definition is given in the frontier section of the documentation);
\item compute the instantaneous and time-cumulative flux of the wave function through this frontier;
\item compute the mass and charge yields associated with this flux;
\item compute the total energy and norm of the system as a function of time;
\item output the evolution of the wave function at any dumped time in a user-specified format;
\item write some global information about the time evolution.
\end{itemize}

\section{Benchmarks}
\label{sec:benchmarks}
In this section, we investigate the convergence properties and overall performance of the numerical methods implemented in \theCode-2.0. This analysis is performed based on the results of three benchmark runs. The first one simulates an oscillating system in a 2-dimensional harmonic oscillator and its exact solution is known analytically. The other two consist in realistic calculations of the low-energy fission yields of \Fm256 and \Pu240 in a $(Q_{20},Q_{30})$ collective space.

\subsection{Harmonic oscillator in two dimensions}
\label{subsec:harmonicOscillator}

\subsubsection{Description of the benchmark}
We study the dynamics of a two-dimensional quantum system in an isotropic harmonic oscillator (HO) potential. The existence of an analytic solution for such motion gives us the opportunity to rigorously test the implementation and compare the convergence speed of different numerical schemes. The HO Hamiltonian reads
\begin{equation}
\hat{H}(\qvec) = -\frac{\hbar^2}{2m}\gras{\nabla}^{2} + \frac{1}{2} m\omega^{2} \qvec^{2}.
\end{equation}
In the context of \theCode-2.0, it is implemented by choosing a diagonal, space-independent inertia tensor inversely proportional to the mass $m$, $B_{kl}(\qvec)=\delta_{kl}/m$. The metric $\gamma(\qvec)$ is set to a non-zero constant over the whole domain: from (\ref{eq:Hcoll}), we see that this choice implies the metric plays no role in the dynamics. The values of the oscillator constants are reported in table~\ref{tab:hoCharacteristics}.

\begin{table}[!ht]
\begin{center}
 \begin{tabular}{ccccc}
 \hline
 m   & $\omega$ &$\Omega$ & $t_\text{max}$ & $\delta t$  \\
 \hline
 1.3 & 0.8      & [-20;20]$\displaystyle^2$ & 16 & $5.10^{-4}$ \\
  \hline
 \end{tabular}
 \caption{Characteristics of the 2-dimensional harmonic oscillator used for this benchmark.}
 \label{tab:hoCharacteristics}
 \end{center}
\end{table}

Calculations are performed in a domain $\Omega=[-20;20]^2$ with Dirichlet conditions imposed at the boundaries $g(\qvec \in \partial \Omega) = 0$. With these parameters, we find that the first two eigenstates of the HO verify
\begin{equation}
\forall \qvec \in \partial \Omega: \quad
 \frac{|g(\qvec)|}{\text{max}\{ \, |g(\qvec)|\, \} }_{\qvec \in \Omega } < 2.10^{-13}
\end{equation}
along the boundaries of the domain. Therefore, the numerical error coming from the finite size of the domain of resolution $\Omega$ is  negligible compared with the other sources of error under study. Noting $\qvec = (q,q')$, the initial wave function is set to
\begin{equation}
 g(q,q',t=0)= \operatorname{exp}\left(-\frac{\tilde{q}^2+\tilde{q}'^2}{2}\right) (1+\tilde{q}),
\end{equation}
where $\tilde{q}$ and $\tilde{q}'$ are the reduced coordinates associated with $q$ and $q'$ through the
relation
\begin{equation}
 \tilde{q} = \sqrt{\frac{m\omega}{\hbar}} q.
\end{equation}
Starting from this state, the analytic time-dependent solution reads
\begin{equation}
  g(q,q',t)= \operatorname{exp}\left(-\frac{\tilde{q}^2+\tilde{q}'^2}{2}\right) e^{-i\omega t}
  \left(1+\tilde{q} e^{-i \omega t}\right).
\end{equation}
The system oscillates from one side of the line $q=0$ to the other side with a period 2$\pi/\omega\simeq 31.4$ a.u. In this benchmark, we compute the evolution of the system up a time $t_\text{max} = 16$ a.u., which roughly corresponds to half the period of oscillation. The time step is set to $dt =5.10^{-4}$ but similar calculations were also performed at $dt = 1.10^{-4}$ and lead to the same conclusions.
We define an arbitrary frontier crossing the 2D box from the bottom to the top. As discussed in \cite{regnier_felix-1.0:_2016}, the flux through any segment of this frontier is known analytically and can be compared with the numerical solution.
We performed a series of calculations with a mesh composed of rectangles, a polynomial basis of degree $d\in[1;4]$ and the spatial steps $h_x = h_y = h =$ 4.0, 2.0, 1.0, 0.5, 0.25, both with and without the spectral approximation.

\subsubsection{Unitarity}
\begin{figure}[!ht]
  \begin{center}
  \includegraphics[width=0.50\textwidth]{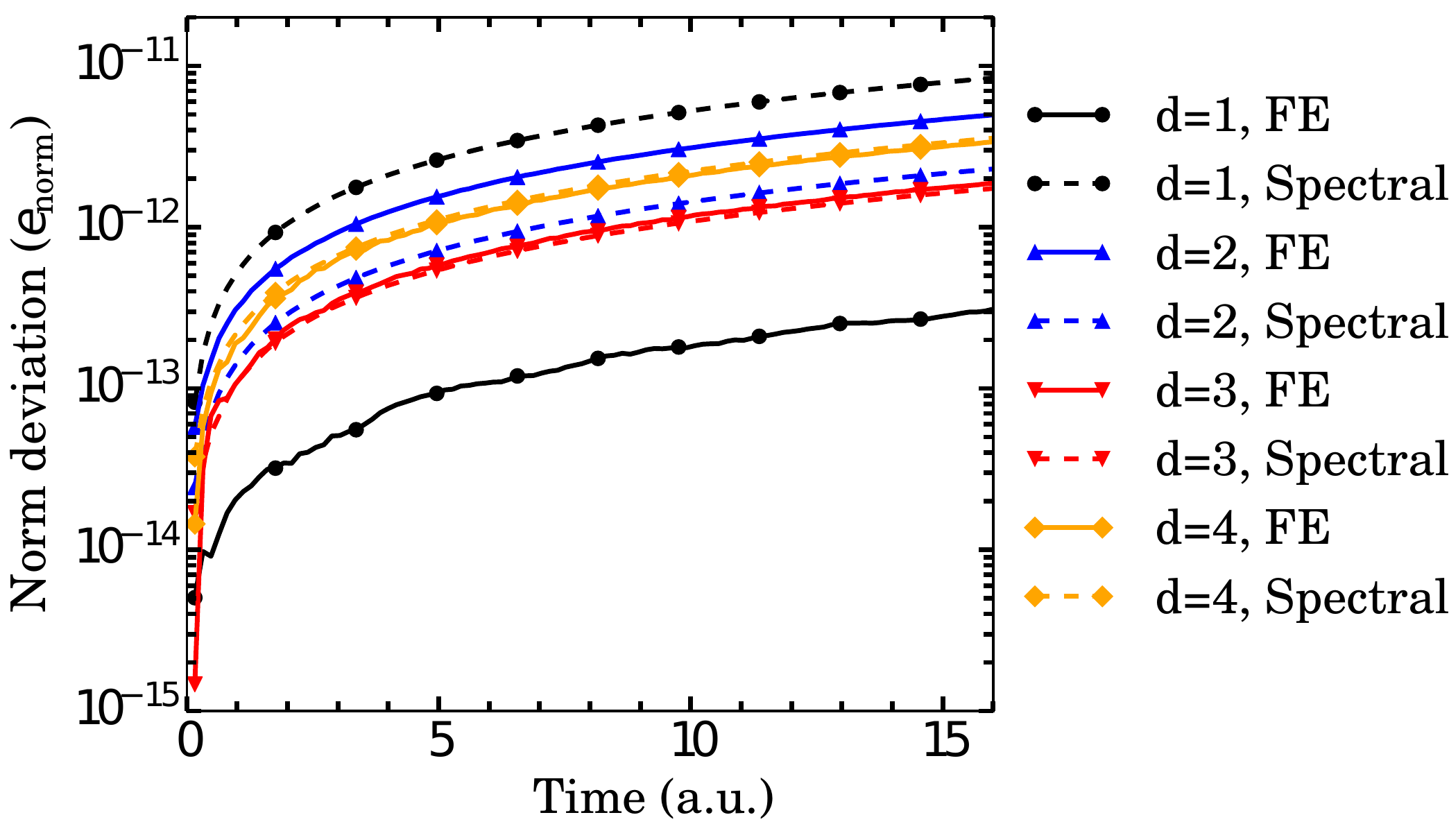}
  \caption{Deviation of the norm $e_\text{norm}$ as a function of time for a periodic motion in a HO potential. Results obtained with a mesh parameter $h=1.0$ and for a finite element basis of degree $d\in[1,4]$. 'FE' denotes a finite element calculation without spectral approximation.}
  \label{fig:err_norm}
  \end{center}
\end{figure}

We first test the unitarity of the time propagator by checking the stability of the norm of the numerical solution as a function of time. The norm deviation is computed as
\begin{equation}
 e_\text{norm}(t) = \left| \frac{ ||g(t)||_2 - ||g(t=0)||_2 }{|| g(t=0) ||_2} \right|.
\end{equation}
Over all the parameter space explored, this deviation stays below $10^{-11}$. The figure~\ref{fig:err_norm} shows the results obtained with a mesh parameter $h=1.0$. It shows the accumulation of a small residual error at each time step. Such a deviation from the expected constant behavior typically comes from round-off errors appearing during the Arnoldi process applied at each time step (cf. section~\ref{sec:krylov}). In practical calculations, this type of error is by far less important that the one associated with the spatial and temporal convergence.

\subsubsection{Spatial convergence}
\begin{figure}[!ht]
  \begin{center}
  \includegraphics[width=0.50\textwidth]{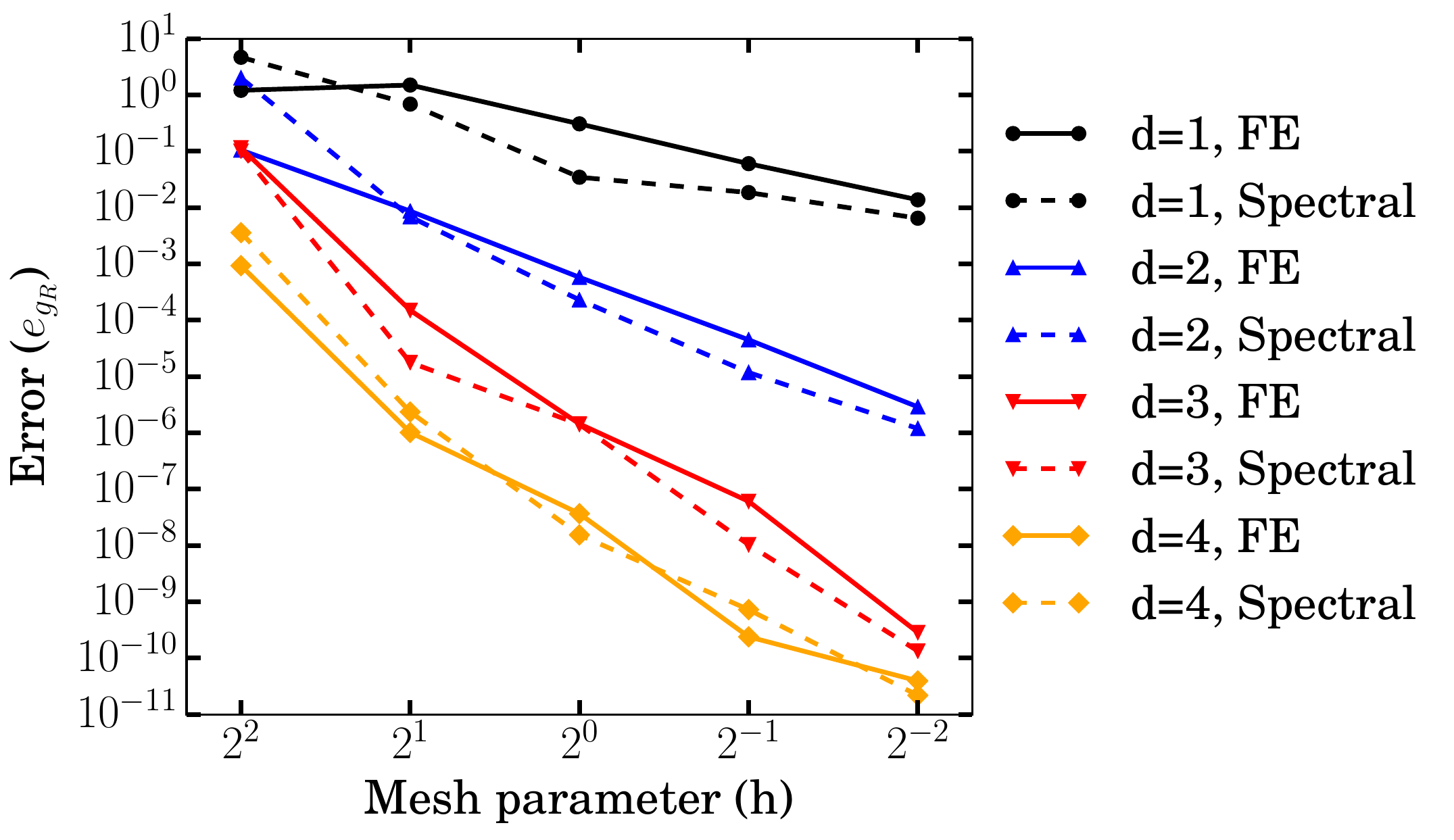}
  \includegraphics[width=0.50\textwidth]{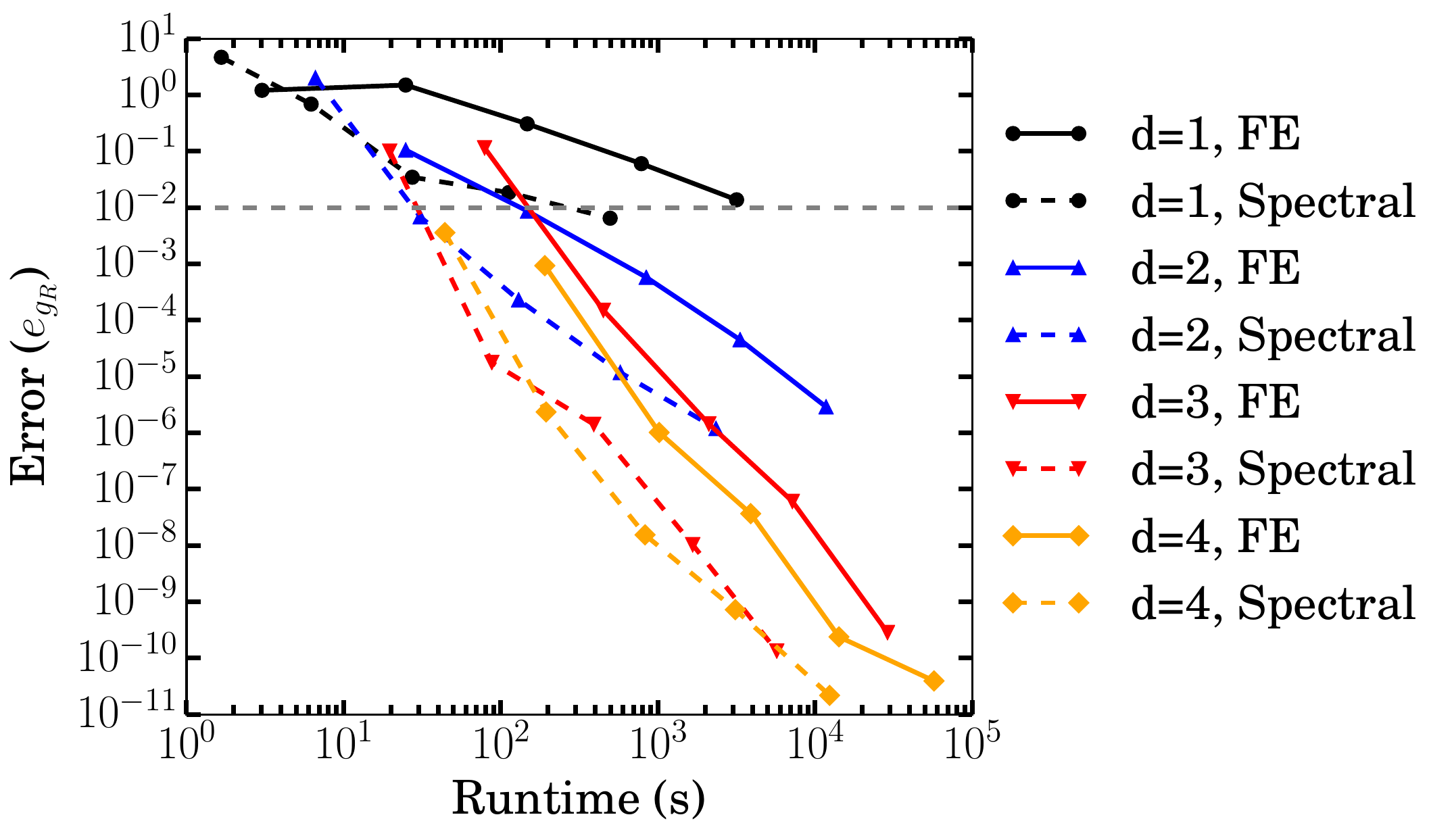}
  \caption{Numerical error on the real part of the collective wave function $e_{\mathrm{g_R}}$ solution for the 2-dimensional HO benchmark. The same error is plotted as a function of the mesh parameter $h$ (top) and runtime (bottom). The calculations are repeated for different values of the degree $d$ of the polynomial basis and for two quadrature schemes. 'FE' denotes a finite element calculation without spectral approximation.}
  \label{fig:err_gR}
  \end{center}
\end{figure}

We now turn to the convergence of the wave function on the frontier at the end of the dynamics evolution. To do so, we define the error $e_{\mathrm{g_R}}$ on the real part of the collective wave function,
\begin{equation}
 e_{\mathrm{g_R}}=  \frac{ || \mathfrak{Re}(g_{\mathrm{num}}(t_\text{max})) - \mathfrak{Re}(g_{\mathrm{theo}}(t_\text{max}))||_\infty }{|| \mathfrak{Re}(g_{\mathrm{theo}}(t_\text{max})) ||_\infty}.
\end{equation}
The infinite norm is computed from the vertices of the mesh present in the frontier. The top part of figure~\ref{fig:err_gR} summarizes the results obtained for the various spatial schemes implemented. All numerical schemes converge toward the analytic solution with an exponential rate typical of the finite element method. The slowing down of convergence at the last points of the $d=4$ curve may be attributed to other sources of error that can no longer be neglected at this level of precision such as, e.g., deviation of the norm.
All dynamical calculations were performed with one thread on the same Intel(R) Xeon(R) CPU E5-2660 v4 @ 2.00GHz processor. The bottom part of figure~\ref{fig:err_gR} shows the convergence as a function of runtime. The most efficient numerical scheme depends on the target precision. For a precision characterized by $e_{\mathrm{g_R}} > 10^{-2}$, the degree 2 scheme is the fastest; for a precision with $e_{\mathrm{g_R}} < 10^{-3}$, the degree 3 and 4 methods become more efficient. Note that whatever the chosen degree, the spectral element method is nearly an order of magnitude faster than the standard finite elements.

\begin{figure}[!ht]
  \begin{center}
  \includegraphics[width=0.50\textwidth]{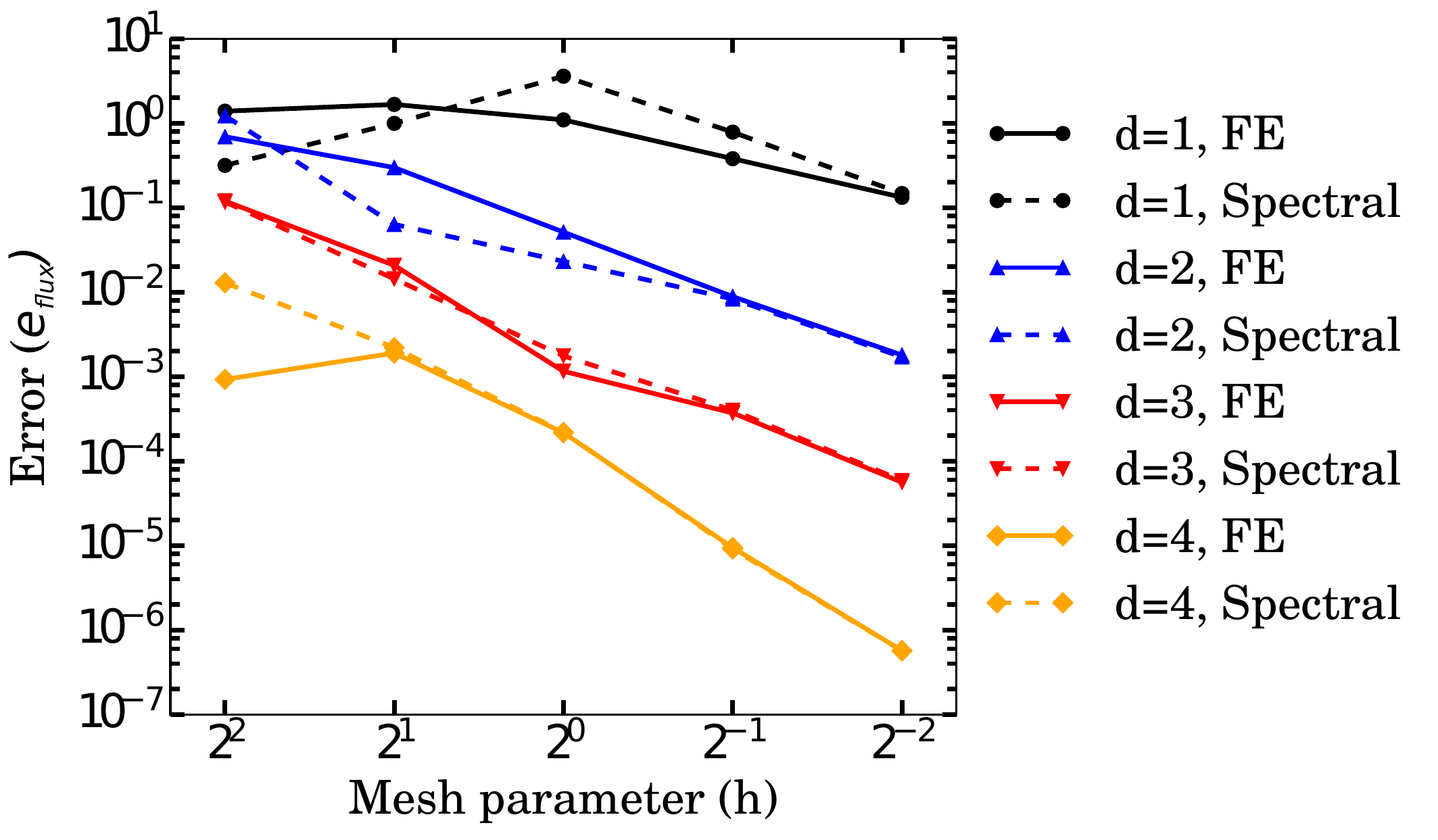}
  \caption{Numerical error of the flux $e_\text{flux}$ as a function of the mesh parameter $h$. The flux is obtained from the solutions of a 2-dimensional HO benchmark obtained with different values of the degree $d$ of the polynomial basis and two quadrature schemes. 'FE' denotes a finite element calculation without spectral approximation}
  \label{fig:err_flux}
  \end{center}
\end{figure}

The convergence of the flux through each element of the frontier is also explored. We define a similar error estimator $e_\text{flux}$ for the flux
\begin{equation}
 e_\text{flux}=  \frac{ || J_{\mathrm{num}}(t_\text{max}) - J_{\mathrm{theo}}(t_\text{max})||_\infty }{|| J_{\mathrm{theo}}(t_\text{max}) ||_\infty},
\end{equation}
where $J(t)$ is the instantaneous flux through each elementary face of the frontier.
The figure~\ref{fig:err_flux} shows the convergence of the flux for all the methods explored. The error also vanishes exponentially but with a slower rate compared to $e_{\mathrm{g_R}}$. As already noticed in~\cite{regnier_felix-1.0:_2016}, this rate is actually limited by the convergence of the derivative of the collective wave function. Our estimate of the derivative could be improved in future versions by leveraging derivative recovery methods as discussed in~\cite{zhang_finite_2011} and references therein. Although some local differences are observed between the spectral and conventional methods, the spectral approximation does not degrade significantly the convergence of the flux.

\subsection{Low energy fission of Fermium 256}
\label{sec:fissionFm256}

\subsubsection{Description of the benchmark}

This benchmark describes a realistic case of low energy fission of \Fm256. The collective Hamiltonian is computed based on the D1S parametrization of the Gogny energy density functional. The potential energy and inertia landscape consists of approximately 18,000 points in the $(Q_{20},Q_{30})$ deformation space ranging from 0 to 450 b along $Q_{20}$ and 0 to $100$ b$^{3/2}$ along $Q_{30}$. The inertia tensor is obtained from the cranking approximation of the GCM inertia and the metric field is obtained as described in~\cite{regnier_fission_2016}. Mesh points are regularly spaced and the grid is characterized by the steps $h_{20} = 2$ b, $h_{30} =1$ b$^{3/2}$. Each point is a result of a constrained HFB calculation, and the complete map is generated using the procedure described in~\cite{regnier_fission_2016} to avoid the presence of local minima.

We first process this original dataset with the {\tt flx-setup} tool to generate the finite element basis and fields. At this stage, we keep only the points with neck $\langle\hat{Q}_N\rangle > 7.0$ to avoid discontinuity problems near scission. The deformation domain is augmented with a band of width 30 in barn units. In this band: (i) the absorption field is activated according to (\ref{eq:absorption}) with the parameters $r=10$ and $w=30$, (ii) the potential energy is set to decrease as a function of the distance to the fission valley, with a slope of $4.10^{-2}$ (MeV per distance unit), while all other fields are extrapolated in a standard way from their values in the original deformation domain. Within a distance 50 (in barn units) to the ground-state, we perform one step of h-refinement for cells with an energy lower than 40 MeV above the ground state. This preprocessing is repeated for different choices of finite elements and mesh parameters.

The initial collective state is defined as an eigenstate of a 2-dimensional harmonic oscillator located in the first potential well. The oscillator is centered near the minimum of the first potential well, at $q_0 = (33\, \mathrm{b}, 0\, \mathrm{b}^{3/2})$. The parameters of the oscillator are shown in table~\ref{tab:fmHOParam}. 

\begin{table}[!ht]
\begin{center}
 \begin{tabular}{cccc}
 \hline
 $\omega_{20}$ (zs$^{-1}$) &$\omega_{30}$ (zs$^{-1}$) & $m_{20}$ & $m_{30}$  \\
 \hline
 1.0 & 8.0  & 3.30 & 0.8  \\
  \hline
 \end{tabular}
 \caption{Characteristics of the harmonic oscillator used to compute the initial state}
 \label{tab:fmHOParam}
 \end{center}
\end{table}

They are chosen to mimic the typical wave lengths and energy of a realistic case with an initial energy lying around 1 MeV above the first saddle point. With this choice for the initial state, the initial collective wave function has the analytic expression
\begin{align}
 \mathfrak{Re}(g(q_{20}, q_{30}, t=0)) & = 
 \operatorname{exp}\left(\frac{\tilde{q}_{20}^2 + \tilde{q}_{30}^2}{2}\right)
 H_2(\tilde{q}_{20}) H_1(\tilde{q}_{30}), \nonumber \\
 \mathfrak{Im}(g(q_{20}, q_{30}, t=0)) & = 0,
\end{align}
where $H_k$ is the k'th Hermite polynomial, and $\tilde{q}_{i0} = \sqrt{m_{i0}\omega_{i0}/\hbar} \ q_{i0}$, $i=2,3$. Using this analytic definition prevents us from introducing additional numerical errors during the calculation of the initial state.

Once the initial state is defined, all relevant matrices are computed with different quadratures and the dynamics is solved for various time steps. We set the total evolution time to $20$ zs (1 zs = $10^{-21}$ s). Finally, we compute the fission mass yields on a frontier defined as the isoline $\langle\hat{Q}_N\rangle = 8.0$. The final yields are obtained after convoluting with a Gaussian of width $\sigma=4.0$.

\subsection{Convergence with the frontier}
\label{sec:frontconv}
Recall that the fission yields are obtained from the flux of the collective wave packet through a hyper-surface called the frontier. Therefore, we expect the fission yields to be sensitive both to the criterion used to define this frontier and to the finite element representation of the mesh at/near the frontier. In the previous release, the frontier was defined as the union of a set of faces of the cells present in the mesh used to solve the dynamics. This version offers more flexibility: a frontier is defined as a set of cell faces of a distinct mesh $M_f$ that can be tuned independently of the mesh used by the solver. In particular, the user can define a new finite element basis $B_f$ specific to the mesh $M_f$. To determine the flux through the frontier with the new mesh $M_f$, the code will
\begin{enumerate}
 \item evaluate the solution of the dynamics at each node of the new basis $B_f$;
 \item compute the flux of the reconstructed wave function through the frontier defined on $M_f$.
\end{enumerate}
In this benchmark, the frontier is defined as the line of cell faces of $M_f$ which best approach the isoline $\langle\hat{Q}_N\rangle = 8.0$ (from the higher values). If $M_f$ is different from the solver mesh, two features emerge: (i) the change of finite element basis may lead to a loss of numerical precision in the evaluation of $g(\qvec,t)$ along the frontier; (ii) on the other hand, if $M_f$ is more refined than the original mesh, the discretized version of the frontier will better match the requested isoline. This may significantly improve the ``geometrical'' precision with which the frontier is defined.

To study these effects, we first performed a reference dynamical calculation hereafter labeled as {\it hcube3}. As its name indicates, the reference calculation is based on a mesh of rectangular cells with mesh parameters $h_{20} = h = 3$ b and $h_{30}= h /3 = 1$ b$^{3/2}$. We used a polynomial basis of degree $d=3$ and performed the dynamics without spectral approximation. The selected time step is $dt = 5.10^{-5}$ zs. The frontier for this reference calculation uses mesh parameters four times smaller than the one used for the propagation, hence $h_f = 0.75$.

From the solution of this reference calculation, two series of frontiers characterized by different meshes $M_f$ and bases $B_f$ are tested. The first one is obtained on a simplex mesh with a finite element basis of degree 2 ({\it simplex2$_f$}), whereas another set is obtained on a rectangular cell mesh with a finite element basis of degree 3 ({\it hcube3$_f$}). In each case, we vary the mesh step $h_f$ used to compute the frontier. The impact of the frontier discretization on the fission yields is quantified with the deviation $e_Y$ from the reference calculation,
\begin{equation}
e_Y = || Y(A) - Y_{\text{ref}}(A) ||_{\infty}.
\end{equation}
where $Y(A)$ are the yields obtained (normalized to 200\%) and $Y_\text{ref}(A)$ are the yields of the reference calculation.

\begin{table}[ht]
\begin{center}
  \begin{tabular}{c|ll}
  \hline
$h_f$    &  {\it simplex2$_f$}  &  {\it hcube3$_f$}   \\
\hline
3    & 0.226  & 0.238 \\
2.12 & 0.149  & 0.130 \\
1.5  & 0.041  & 0.069 \\
1.06 & 0.027  & 0.034 \\
0.75 & 0.022  & {\bf 0.000} \\
  \hline
  \end{tabular}
\caption{Deviation $e_Y$ of the yields as a function of the mesh parameter $h_f$ used to compute the frontier for two different mesh types. The calculation of reference is highlighted in boldface.}
\label{tab:frtconv}
\end{center}
\end{table}

The results are reported in table~\ref{tab:frtconv}. 
They present a clear convergence as the frontier cells decrease in size. Demanding a relative precision on the yields of less than 5\% requires a mesh parameter of the order of 1 b for the definition of the frontier. Performing the full time evolution of the fissioning system with such a level of mesh refinement would represent a costly and unnecessary task (see section~\ref{sec:waveconv}). A better method consists in performing the time evolution with a coarser spatial scheme that is sufficient to ensure the convergence of the evolution of the collective wave packet, and then compute the flux on a refined mesh ($h_f\simeq 1$).

\subsubsection{Convergence of the wave packet propagation}
\label{sec:waveconv}

This section focuses on the convergence of the wave packet itself at the beginning and the end of its propagation. We perform series of calculations with different meshes, finite element bases and quadratures, as well as different time step values. The mesh parameters are taken as $h_{20}= 3 h_{30} = h$ with $h = 8.49, 6, 4.24, 3$ (in barn units). Each element of this series decreases the mesh parameter by a factor $\sqrt{2}$ and therefore increases the number of nodes involved by a factor of 2. We consider the following values for the time step, $dt = 4.10^{-4}, 2.10^{-4}, 1.10^{-4}, 5.10^{-5}$ zs. Finally, we select a sample of a few meaningful finite element bases:
\begin{itemize}
 \item \textit{simplex2}: A degree 2 finite element basis using simplex cells. The nodes are located regularly inside the element, and matrix elements are computed with a Gauss-Legendre quadrature.
 \item \textit{hcube3}: A degree 3 finite element basis using rectangular cells. The nodes are located at the Gauss-Lobatto quadrature points of order $d+1$. The matrix elements are computed exactly with a Gauss-Legendre quadrature.
 \item \textit{spectral2(3,4)}: A degree $d=2$ (3,4) element basis using rectangular cells. The nodes are located at the Gauss-Lobatto quadrature points of order $d+1$. The matrix elements of $H$ are computed with a Gauss-Legendre quadrature whereas $M$ is obtained with the spectral approximation (Gauss-Lobatto quadrature of
 order $d+1$).
\end{itemize}
To avoid numerical uncertainties coming from the definition of the frontier itself, see section \ref{sec:frontconv}, the flux and fission yields are computed from the same frontier in all this section. We adopted the frontier of the reference calculation detailed in section~\ref{sec:frontconv} ($h_f=0.75$). Recall that the reference calculation used to assess numerical convergence corresponds to the \textit{hcube3} method with $h=3$ and $dt = 5.10^{-5}$ zs.

\begin{table}[!h]
\centering
  \begin{tabular}{l|cccc}
  \hline
  Method/$h$ &     8.49 & 6 & 4.24 & 3 \\ 
  \hline
  simplex2   & -1883.264 & -1883.376 & -1883.400 & -1883.403  \\
  hcube3     & -1883.409 & -1883.406 & -1883.401 & {\bf -1883.405}  \\
  spectral2  & -1856.569 & -1875.890 & -1881.475 & -1882.916  \\
  spectral3  & -1882.453 & -1883.324 & -1883.381 & -1883.402  \\
  spectral4  & -1883.382 & -1883.402 & -1883.405 & -1883.405  \\
  \hline
  \end{tabular}
\caption{Energy of the initial state $g(\qvec, t=0)$ in MeV.  The calculation of reference is highlighted in boldface.}
\label{tab:initEnergy}
\end{table}

We report in Table~\ref{tab:initEnergy} the energy of the initial state after discretization on the finite element basis. The convergence of this energy with mesh size and type of finite element basis already gives us information about the quality of the spatial discretization in the vicinity of the ground-state. Most methods converge to -1883.405 $\pm$ 5 keV, which is about 13.25 MeV above the minimum of the first potential well and 0.87 MeV above the first saddle point. The \textit{spectral2} scheme seems to converge too slowly to reach a $\pm$ 5 keV accuracy within the range of mesh parameters explored. 

\begin{table}[!ht]
\begin{center}
 \begin{tabular}{c|c|cccc}
 \hline
Method       & $dt$ (zs) / $h$ (b) & 8.49 & 6 & 4.24 & 3    \\
 \hline
simplex2  & $4.10^{-4}$& 19.56 & 23.44 & 24.34 & 25.11   \\
          & $2.10^{-4}$& 19.60 & 23.63 & 24.34 & 25.03   \\
          & $1.10^{-4}$& 19.59 & 23.61 & 24.34 & 25.03   \\
          & $5.10^{-5}$& 19.59 & 23.61 & 24.34 & 25.03   \\
 \hline                                             
hcube3    & $4.10^{-4}$& 24.42 & 25.00 & 12.40 &           \\
          & $2.10^{-4}$& 24.40 & 24.81 & 24.97 & 11.96     \\
          & $1.10^{-4}$& 24.42 & 24.80 & 24.82 & 24.92     \\
          & $5.10^{-5}$& 24.42 & 24.80 & 24.83 & {\bf 24.87}     \\
\hline                                             
spectral2 & $4.10^{-4}$& 09.04 & 13.59 & 20.98 & 22.90    \\
          & $2.10^{-4}$& 09.02 & 13.40 & 21.18 & 23.69    \\
          & $1.10^{-4}$& 09.09 & 13.35 & 21.11 & 23.70    \\
          & $5.10^{-5}$& 09.05 & 13.35 & 21.12 & 23.70    \\
\hline                  
spectral3 & $4.10^{-4}$& 15.42 & 23.84 & 25.00 & 10.78  \\
          & $2.10^{-4}$& 15.42 & 23.98 & 24.55 & 25.11  \\
          & $1.10^{-4}$& 15.44 & 23.95 & 24.58 & 24.87  \\
          & $5.10^{-5}$& 15.44 & 23.95 & 24.58 & 24.87  \\
\hline                 
spectral4 & $4.10^{-4}$& 24.98 & 24.89 & 11.16 &       \\
          & $2.10^{-4}$& 24.86 & 24.83 & 23.95 & 12.02  \\   
          & $1.10^{-4}$& 24.84 & 24.86 & 24.91 & 10.79  \\    
          & $5.10^{-5}$& 24.84 & 24.86 & 24.92 & 24.90  \\    
\hline          
 \end{tabular}
 \caption{Total flux crossing the frontier between 0 and 20 zs as a function of the time and space resolutions. The total flux is expressed in \% of the initial collective wave packet norm. The parameters $dt$ and $h$ are expressed in zs ($10^{-21}s$) and barn units respectively. The calculation of reference is highlighted in boldface.}
 \label{tab:totalFlux}
 \end{center}
\end{table}

Table~\ref{tab:totalFlux} shows the total cumulated flux that crossed the frontier during the time range $[0, 20]$ (zs). We could not reach convergence for two of the space/time discretization schemes studied here when the time-step was too large. The \textit{simplex} and \textit{hcube} methods converge toward a value of $24.9 \pm 0.1$, meaning that nearly 25\% of the collective wave packet crossed the frontier during the simulation time. The remaining numerical noise of $\pm 0.1$ represents less than 0.5\% of the total flux value. Once again, the \textit{spectral2} method converges more slowly than the other schemes. 
%
%
Let us emphasize that it is necessary to lower the time step in order to reach convergence when $h$ decreases. This behavior can be related to the Krylov approximation of the propagator: decreasing $h$ induces an increase of the dimension of the Hamiltonian matrix in (\ref{eq:spaceDiscretizedEvolution}). As the dimension for the Krylov space is kept constant, it thus becomes necessary to lower the time step to approximate the exponential with the same precision.

\begin{table}[!ht]
\begin{center}
 \begin{tabular}{c|c|cccc}
 \hline
Method &      $d t$ (zs) / h (b) & 8.49 & 6 & 4.24 & 3     \\
\hline
simplex2  & $4.10^{-4}$& 0.302 & 0.120 & 0.064 & 0.035     \\
          & $2.10^{-4}$& 0.325 & 0.090 & 0.105 & 0.019     \\
          & $1.10^{-4}$& 0.326 & 0.098 & 0.106 & 0.019     \\
          & $5.10^{-5}$& 0.326 & 0.097 & 0.106 & 0.019     \\
 \hline
hcube3    & $4.10^{-4}$& 0.057 & 0.119 & 5.677 &       \\
          & $2.10^{-4}$& 0.101 & 0.032 & 0.038 & 5.669      \\
          & $1.10^{-4}$& 0.097 & 0.031 & 0.007 & 0.014      \\
          & $5.10^{-5}$& 0.097 & 0.031 & 0.007 & {\bf 0.000}      \\
\hline
spectral2 & $4.10^{-4}$& 0.943 & 0.587 & 0.227 & 0.366     \\
          & $2.10^{-4}$& 0.937 & 0.633 & 0.319 & 0.369    \\
          & $1.10^{-4}$& 0.925 & 0.648 & 0.301 & 0.371    \\
          & $5.10^{-5}$& 0.932 & 0.647 & 0.301 & 0.371    \\
\hline
spectral3 & $4.10^{-4}$& 0.809 & 0.116 & 0.135 & 5.652    \\
          & $2.10^{-4}$& 0.812 & 0.155 & 0.058 & 0.053    \\
          & $1.10^{-4}$& 0.821 & 0.158 & 0.060 & 0.012   \\
          & $5.10^{-5}$& 0.822 & 0.159 & 0.060 & 0.012   \\
\hline          
spectral4 & $4.10^{-4}$& 0.088 & 0.079 & 5.700 &  \\
          & $2.10^{-4}$& 0.116 & 0.014 & 0.035 & 5.662  \\   
          & $1.10^{-4}$& 0.120 & 0.011 & 0.003 & 0.233  \\    
          & $5.10^{-5}$& 0.120 & 0.011 & 0.004 & 0.003  \\    
\hline
 \end{tabular}
 \caption{Deviation $e_Y$ as a function of time and space resolution and for different numerical schemes. Yields are normalized to 200$\%$. The parameters $dt$ and $h$ are expressed in zs ($10^{-21}s$) and barn units respectively. The calculation of reference is highlighted in boldface.}
 \label{tab:yieldsDiff}
 \end{center}
\end{table}

We now compare the fission yields obtained for this set of calculations. To do so, we use the yields deviation $e_Y$ introduced in section~\ref{sec:frontconv} along with the same reference yields (\textit{hcube3} with $h=3$, $dt=5.10^{-5}$ zs). Table~\ref{tab:yieldsDiff} summarizes the results. Except for the \textit{spectral2} scheme, all methods converge within $e_Y<0.02$ (\%) compared to the reference calculation. For the majority of possible mass splits, the value of the fission yields is typically in the range $1\%<Y(A)<7\%$. The \textit{spectral2} method is clearly inadequate to reach such a level of precision. Note that by comparing \textit{hcube3} with \textit{spectral3}, we directly measure the effect of the spectral approximation.

\begin{figure}[!ht]
  \begin{center}
  \includegraphics[width=0.450\textwidth]{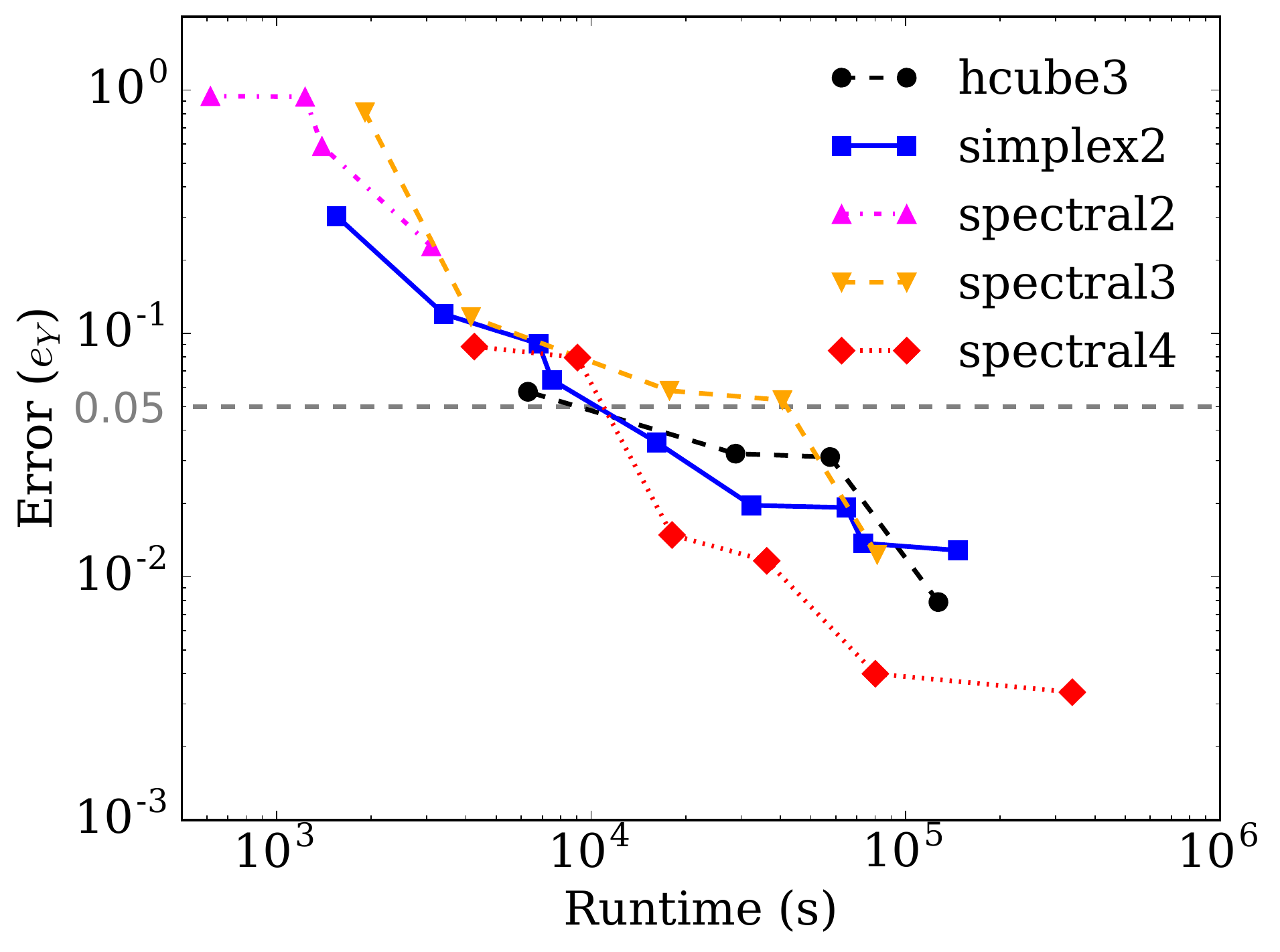}
  \caption{Best runtime obtained with different numerical schemes, mesh parameters and time steps.}
  \label{fig:time_bench}
  \end{center}
\end{figure}

Finally, figure~\ref{fig:time_bench} shows this benchmark in terms of calculation runtimes. Here, the runtime is  the execution time for the collective wave packet propagation on one thread of a Intel(R) Xeon(R) CPU E3-1271 v3 @ 3.60GHz processor. Each point in the figure corresponds to a single calculation of table~\ref{tab:yieldsDiff}. A calculation is shown only if no other calculation has both a shorter runtime and a better convergence (i.e., a lower $e_Y$). For a convergence $e_Y > 0.05$, the \textit{simplex2} scheme is the most efficient, whereas for better precision the most efficient scheme is clearly \textit{spectral4}. This last one is about 2 times faster than the others for a target precision of $e_Y=10^{-2}$.

\subsubsection{Multi-threading}
The most CPU-intensive part of \theCode-2.0 is the repetitive linear algebra operations performed during time propagation. These operations are outsourced to the Eigen-3.3.2 library~\cite{guennebaud_eigen_2010} which offers support for multi-threading. As a consequence, the code can benefit from a speedup with the number of cores even though the time-loop of the TDGCM+GOA solver does not implement any explicit shared memory parallelism. We analyze this potential speedup for four spatial schemes having similar numbers of nodes, namely, \textit{hcube3} with $h=4.24$ ($2.2\, 10^5$ nodes), \textit{simplex2} with $h=2.12$ ($3\, 10^5$ nodes), \textit{spectral3} with $h=4.24$ ($2.2\, 10^5$ nodes) and \textit{spectral4} with $h=6$ ($1.9\, 10^5$ nodes). For each scheme, the benchmark consists in repeating 10 times a calculation with 100 time iterations ($dt = 5.10^{-5}$ zs). The speedup achieved with $n\geq1$ threads is calculated as the average over the 10 batches of the time spent in the time loop divided by the same quantity obtained with one thread. Calculation were performed on a Intel(R) Xeon(R) CPU E3-1271 v3 @ 3.60GHz processor.

\begin{figure}[!ht]
  \begin{center}
  \includegraphics[width=0.450\textwidth]{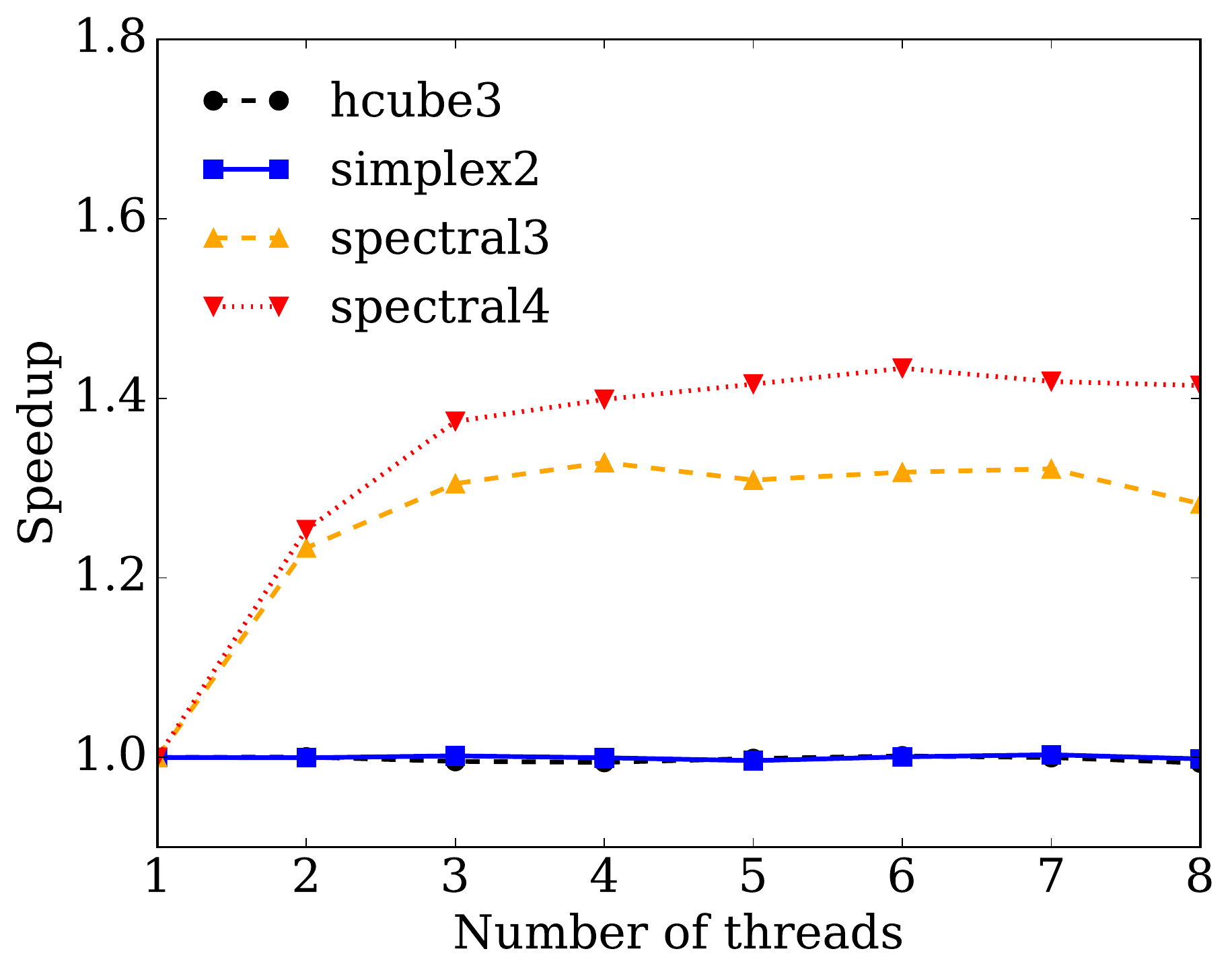}
  \caption{Speedup obtained by increasing the number of threads used to solve the wave packet propagation.}
  \label{fig:multithread_bench}
  \end{center}
\end{figure}

The figure~\ref{fig:multithread_bench} shows the speedup as a function of the number of threads used for the calculations. The \textit{hcube3} and \textit{simplex2} schemes do not benefit from any noticeable speedup. These schemes both involve inversions of triangular matrices ($L$ and $L^t$) at each time step, which are not accelerated with additional threads. The \textit{spectral} method is free of this bottleneck, and presents a small enhancement with the number of threads. However, this speedup rapidly saturates around 1.4. We observe that potential gains from multi-threading linear operations are limited compared to the overall solver runtime. We will study how to increase the strong scaling of the code in future releases of the code.

\subsection{Comparison to \theCode-1.0}
\label{sec:felix-1.0}

In this section we compare the performance of \theCode-2.0 with \theCode-1.0 on the calculation of the mass yields for the low-energy neutron-induced fission of \Pu239. This benchmark was already performed with \theCode-1.0 and is extensively presented in~\cite{regnier_felix-1.0:_2016}. Let us briefly recall here that it consists in computing the dynamics of the compound system up to a time 6 zs. 
The initial state is built as a Gaussian wave packet in the collective variables, which is then multiplied by a plane wave in the positive $Q_{20}$ direction. It is characterized by the width of the Gaussian $\sigma_{20} = 23.2$ b, $\sigma_{30}=6$ b$^{3/2}$, the center of the Gaussian $Q_{20} = 30$ b and $Q_{30}=0$ b$^{3/2}$, as well as the momentum of the plane wave $k=0.26$ b$^{-1}$. These numerical values are the same as the ones used in our previous work except for a slight change in $\sigma_{20}$ ($=26$ b in the previous one). This change was needed in order to keep the energy of the initial state at about 500 keV above the fission barrier, since the new version uses a different extrapolation method in the $Q_{20}<0$ area. 

Calculations with \theCode-2.0 are performed using the \textit{simplex2} scheme with the set of mesh parameters explored in the previous paper $h=7.93, 5.95, 4.76, 3.97, 3.40, 2.98, 2.64$ (in b units). We perform this series of calculations with the different time steps $dt=4.10^{-4}, 2.10^{-4}$ zs. Runtimes are determined based on the use of a single thread of a Intel(R) Xeon(R) CPU E3-1271 v3 @ 3.60GHz processor. The error at convergence is estimated with the $e_Y$ deviation, the reference calculation being the one obtained with the smallest mesh parameter and time step for each code version.

\begin{figure}[!h]
  \begin{center}
  \includegraphics[width=0.450\textwidth]{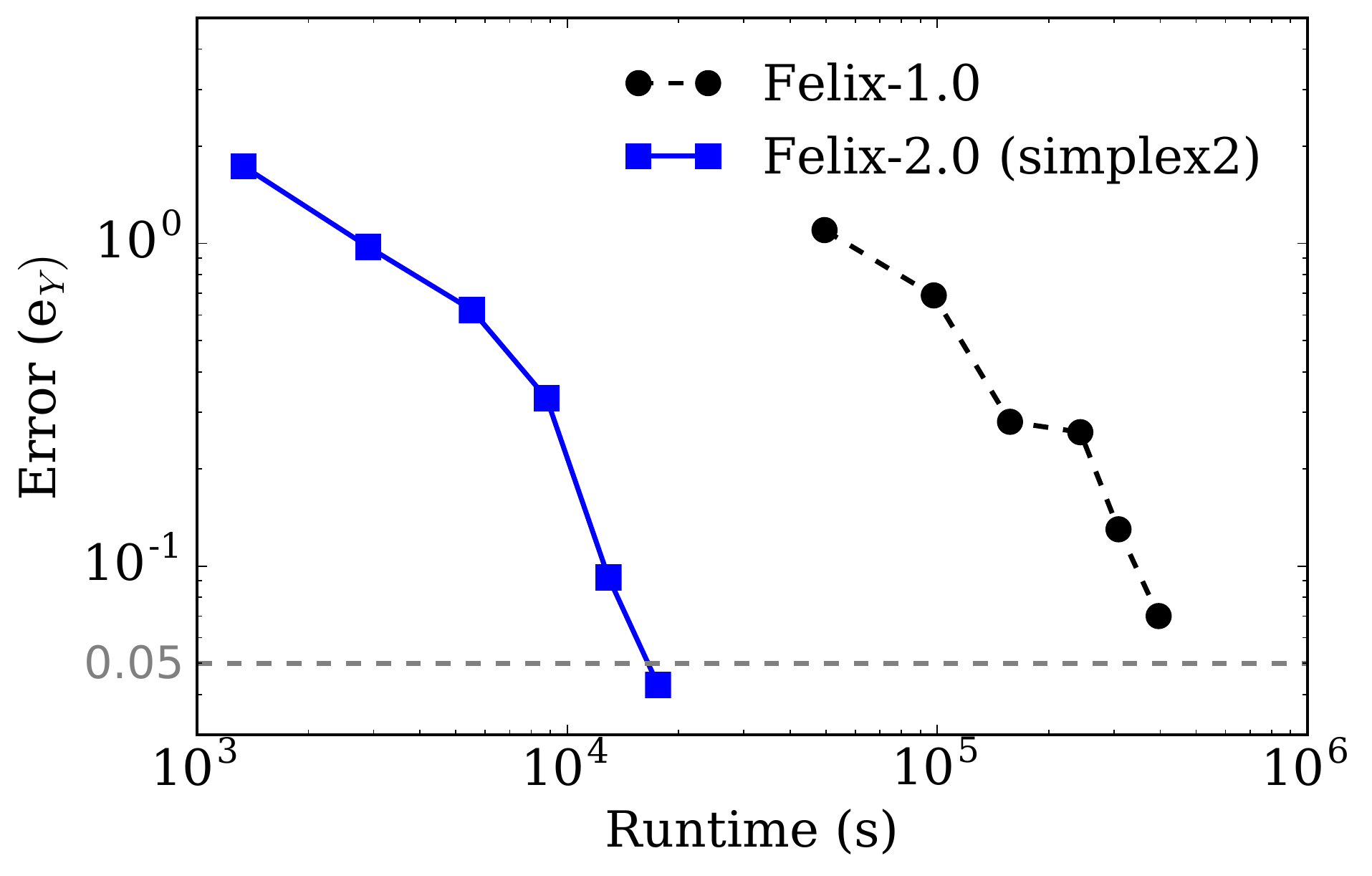}
  \caption{Deviation $e_Y$ for the low energy fission of \Pu240 as a function of runtime and with different mesh parameters. Calculations are performed with the time steps $dt=1.10^{-4}$ and $dt=4.10^{-5}$ zs ($10^{-21}$s) for the version 1.0 and 2.0 of \theCode \ respectively.}
  \label{fig:pu240_perf}
  \end{center}
\end{figure}

The deviation $e_Y$ obtained for two calculations with the same mesh parameter characterizes the convergence in time. With \theCode-1.0, the time steps adopted in our previous work were $dt= 1.10^{-4}$ and $5.10^{-5}$ zs. The convergence in time was typically $|e_Y(1.10^{-3}) - e_Y(5.10^{-4})| = 0.02$. With \theCode-2.0, the convergence with our selected time steps is $|e_Y(4.10^{-4}) - e_Y(2.10^{-4})| = 0.001$. In both cases, the main part of the deviation is coming from the spatial resolution. We compare the deviation obtained with the different mesh parameters as a function of the calculation runtime in figure~\ref{fig:pu240_perf}. For a typical convergence $e_Y=0.05$ \%, we find that \theCode-2.0 is at least 20 times faster than \theCode-1.0. In this benchmark, the increase in performance is mostly caused by the Krylov propagation scheme. Based on the discussion in section~\ref{sec:waveconv}, we can expect that using spectral elements could bring an additional speedup when better precision is required.

\section{Usage of \theCode-2.0}
\label{sec:usage}

The package is composed of the following directories and files:
\begin{itemize}
 \item {\tt README.md, AUTHORS, LICENSE}: contains the basic instructions about how to build, use and distribute this package;
 \item {\tt Makefile}: a standard GNU makefile to build the solver, the tools, the tests, and the documentation;
 \item {\tt src/}: C++ source files of the TDGCM solver and of the tools;
 \item {\tt tools/}: additional C++ and Python source files to handle the inputs and outputs of the TDGCM solver;
 \item {\tt tests/}: a unitary test suite for the package as well as a series of global benchmarks;
 \item {\tt examples/}: a few preset inputs and their corresponding outputs;
 \item {\tt doc/}: documentation of the package in DoxyGen format.
\end{itemize}

The full Felix package depends on several standard Open Source libraries:
\begin{itemize}
 \item The TDGCM solver itself requires BLAS, LAPACK, and a C++ compiler with
OpenMP support;
 \item Generating a local HTML documentation requires DoxyGen-1.8.6 or higher;
 \item In order to build the full set of tools included in this release, the user must also install Xerces-C and the Boost unit test framework. The versions Xerces-C-3.1 and Boost-1.58 have been used during development.
\end{itemize}
This version also uses the Eigen-3.3.2~\cite{guennebaud_eigen_2010} linear algebra library, the Spectra-0.2.0 library for sparse eigen-problems and the Pugixml-1.7 to parse XML files. These components are provided with this release (in directory {\tt src/thirdparty/}) and will be compiled when needed by the present Makefile.

\subsection{Compilation}

As in the previous release, the program is shipped with a Makefile containing a preset configuration assuming  compilation with the GNU Compiler Collection (GCC) on a standard LINUX distribution. The user should expect to have to change this Makefile to match his/her own system configuration. The different components of the package can be compiled separately by typing the following commands:
\begin{itemize}
 \item {\tt make doc}: generate the DoxyGen documentation in the directory {\tt doc/DoxygenDoc}.
 \item {\tt make solver}: build the TDGCM+GOA solver executable. Its name is {\tt flx-solver} and it is located  in a newly created directory {\tt bin/}.
 \item {\tt make tests}: build the executable for the full suite of tests included in the package. The name of the executable is {\tt flx-tests} and is located in {\tt bin/}.
 \item {\tt make tools}: compile all tools and append their respective executable in the directory {\tt bin/}.
\end{itemize}
The package has been successfully compiled with gcc 4.4.7 and higher, as well as the Intel compiler icc 15.0.1.

\subsection{Solver inputs}
A successful execution of the solver requires the following inputs:
\begin{itemize}
 \item    The geometrical definition of the problem along with a finite element discretization of the space,
 \item    the potential $V$, the metric $\gamma$ and the inertia tensor $B$ evaluated at each node of the basis,
 \item    an initial wave function $g(\qvec, t=0)$ evaluated at each node,
 \item    a set of options for the calculation itself (e.g. time step, print rate, etc).
\end{itemize}
Other features and options may additionally be specified by the user.
Most of them are handled through the options set, and some will need an additional input file.
The code reads its input from several files named with a common, user-defined prefix followed by a specific extension, for instance, 
{\tt example-pes.xml}, {\tt example-ini.dat}, etc. The prefix may be a directory such as {\tt path/to/inputs/}.
Here, "example-" is the prefix. 

The main and only mandatory input file for the solver is named {\tt example-pes.xml}. It contains the mesh of the simulation domain $\Omega$, a finite element basis and also the value of several fields at the nodes of this basis. This information is packaged in the form of a unique XML file whose format is detailed in the "Inputs \& Outputs" section of the documentation.
Depending on the user options, the initial state can be defined either by one file named {\tt example-ini.dat} and containing a wave packet $g(\qvec, t=0)$,  or by a directory containing a set of files named {\tt state\_xxx.dat}. In any case, all these files have the same simple format described in the documentation. 

\subsection{Solver options}
In addition to the aforementioned inputs, the user may provide a set of options either directly via the command line or in a file given as the only argument to the executable {\tt flx-solver}. The options determine the inputs and outputs path for the solver as well as some parameters of the dynamics calculation (e.g. time step, print rate, etc). The complete set of options as well as their detailed description is given in the documentation, and a summary is provided when executing {\tt flx-solver --help}.

\subsection{Solver outputs}
All the outputs are printed in a series of files starting with the prefix specified via the options.
After a successful execution of the solver, the following outputs should appear:
\begin{itemize}
 \item {\tt solver.opt}: This is a print of the comprehensive option set used for the run. 
     It could be re-used as an input for another calculation.
 \item {\tt dyn.dat}: This binary file stores the wave function at each dumped iteration.
\end{itemize}
For each dumped iteration, a block of data is printed into this file. One block first contains the iteration number (binary encoded integer) and its associated time (binary encoded double). If the finite element basis contains n nodes, the block will then contain the sequence of the n real parts of the g function value at the nodes (binary encoded double), followed by the sequence of its n imaginary parts (binary encoded double). 

\section{Acknowledgements}
\label{sec:acknowledgement}

This software is based on pugixml library (http://pugixml.org). Pugixml is Copyright (C) 2006-2015 Arseny Kapoulkine. This work was partly performed under the auspices of the US
Department of Energy by the Lawrence Livermore National Laboratory under
Contract DE-AC52-07NA27344. An award of computer time was
provided by the Innovative and Novel Computational Impact on Theory and
Experiment (INCITE) program. This research used resources of the Oak Ridge
Leadership Computing Facility located in the Oak Ridge National Laboratory,
which is supported by the Office of Science of the Department of Energy
under Contract DE-AC05-00OR22725. It also used resources of the National
Energy Research Scientific Computing Center, which is supported by the
Office of Science of the U.S. Department of Energy under Contract
No. DE-AC02-05CH11231. 




\section{References}
\bibliographystyle{elsarticle-num}
\bibliography{cpcFelix2.bib}







\end{document}